\documentclass[12pt]{iopart}
\usepackage[dvips]{graphicx}
\begin{document}
\title{Superheavy nuclei with the vector self-coupling of the $\omega$-meson 
in the relativistic mean-field theory}

\author{A A Saldanha, A R Farhan, M M Sharma}

\address{Physics Department, Kuwait University, 
Kuwait 13060, Kuwait}

\begin{abstract}
We have studied properties and shell structure of the superheavy 
elements from $Z=102$ to $Z=120$ within the framework of the RMF theory. 
The region of study spans nuclides with neutron numbers $N=150-190$.
The Lagrangian model NL-SV1 with the inclusion of the vector self-coupling 
of the $\omega$-meson has been employed in this work. We have performed 
RMF + BCS calculations for an axially deformed configuration of nuclei.
The ground-state binding energies, single-particle properties and quadrupole 
deformation of nuclei have been obtained from the mean-field minimizations. 
Two-neutron separation energies, $Q_\alpha$ values and $\alpha$-decay
half-life have been evaluated. It is shown that a large number of nuclides 
exhibit the phenomenon of shape-coexistence over a significant region of the 
superheavy elements. Shape coexistence of a prolate and an oblate shape
is prevalent in nuclides far below $N=184$, whilst nuclei in the vicinity of
$N=184$ tend to show a shape coexistence between a spherical and an oblate 
shape. The shell structure and 2-neutron separation energies 
obtained with the RMF theory reinforce the neutron number $N=184$ 
as a major magic number. It is shown that the neutron number $N=172$ acts 
akin to a magic number in the deformed region. It is suggested that the 
combination $Z=120$ and $N=172$ has the potential of being a doubly 
magic number in the superheavy region.
\end{abstract}
\submitto{\JPG}
\maketitle
\normalsize

\section{Introduction}
Significant progress has been made in the discovery of new superheavy nuclei 
in the last decade \cite{Muenzenberg.88,Hofmann.00,Hofmann.96,Hofmann.02,
Ginter.03,Oganessian.99,Oganessian.00,Oganess.00,Oganessian.04,Oganess.04}.
Superheavy elements at the extreme end of the periodic table have been 
synthesized in the laboratory.
The synthesis of superheavy nuclei with $Z=110-112$ at
GSI, Darmstadt \cite{Hofmann.96,Hofmann.02} and 
JINR, Dubna \cite{Oganessian.99,Oganessian.00,Oganess.00,Oganessian.04,
Oganess.04} has stimulated the interest in the properties of
superheavy nuclei. Discovery of isotopes of $Z=112$ and $Z=114$ was 
reported in JINR, Dubna and synthesis of $Z=115$ at Berkeley was 
reported some time ago \cite{Oganess.04}. 
Earlier superheavy nuclei could be unambiguously 
identified by their $\alpha$-decay chains leading to already known nuclei. 
However, the decay chains of the newly found superheavy nuclei cannot be 
linked to any nuclei known due to their relatively extreme positions in the
periodic table. Therefore, their identification depends much upon a comparison 
with theoretical models.

Much progress has also been made in theoretical investigations 
of the properties of superheavy nuclei. Theoretical models such as 
macroscopic-microscopic (MM) models have been employed to discern 
and predict properties of superheavy nuclei. At the same time, 
various mean-field approaches have also been employed to investigate basic 
properties in this domain. The prediction of $^{298}_{184}$114 was confirmed
by MM models such as FRDM with Folded Yukawa single-particle potential
\cite{Moeller.95}. For the next doubly-magic nucleus, the available
self-consistent models predict this to the more proton rich $^{292}_{172}$120 
or the even heavier one at $^{310}_{184}$126. Predictions in various models 
tend to diverge primarily due differences in the form of the single-particle 
potentials.

Possibly different extrapolations arise in self-consistent Relativistic
Mean-Field (RMF) theory. Here, the spin-orbit coupling is naturally 
built-in as a consequence of the Dirac-Lorentz structure of nucleons, 
whereas in all non-RMF models, it has to be introduced by hand. 
The non-RMF models tend to overestimate spin-orbit splitting with 
increase in mass number, thus affecting actual predictions of shell 
closures in the superheavy region. Superheavy nuclei with inherently 
large density of single-particle states can be a sensitive 
probe for models of nuclear structure.

Various microscopic approaches such as non-relativistic density-dependent 
Skyrme Hartree-Fock (SHF) theory 
\cite{Cwiok.83,Cwiok.96,buervenich.04,Rutz.97,Bender.99,Bender.01,Rein.02}
and that of MM type 
\cite{Sobicz.01,Muntian.04,Muntian.01,Muntian.03,Muntian.03a} 
are used extensively to investigate the properties and structure of superheavy 
nuclei. In the MM approach, a sum of smooth energy based upon liquid-drop type 
formula on which shell correction is imposed using the method of Strutinsky 
is obtained as the total binding energy of nuclei. The Finite-Range Droplet
Model (FRDM) has achieved a significant success amongst the MM models.
The mass formula FRDM is used to calculate the shell correction energies 
in order to identify the major magic numbers in the region of superheavy 
nuclei. A major proton magic number $Z=114$ and a neutron magic number 
$N=184$ have been predicted by the FRDM.

The MM approach has also been performed by Sobiczewski group 
\cite{Sobicz.01,Muntian.04,Muntian.01,Muntian.03,Muntian.03a}, 
where the Strutinsky shell correction \cite{Stru.67,Stru.68} is used 
for the microscopic component, whereas Yukawa + exponential model 
is used for the macroscopic part. Superheavy nuclei have also been 
studied extensively within the self-consistent
mean-field models of Skyrme Hartree-Fock (SHF) 
type. Comparative studies of SHF and RMF
models have been made. One of the first studies \cite{Lala.96}
performed within RMF theory earlier on was using the force 
NL-SH \cite{SNR.93}. It was shown that neutron number $N=184$ appears 
to be magic for lighter superheavy nuclei whereas its importance 
is reduced for heavier superheavies with $Z>114$. There was also an 
indication of a small deformed shell-closure at $Z=114$ for the 
proton number. In comparison, the FRDM predicted a strong shell 
gap at $Z=114$ in the deformed region.

Extensive shell correction calculations were performed by 
Kruppa et al. \cite{Kruppa.00} for sets of the Skyrme and RMF forces 
in order to identify the magic numbers in the region of superheavy elements 
theoretically. The Skyrme models predict the strongest shell effects 
for $N=184$ and $Z=$ 124, 126 and not at $Z=114$, whereas RMF theory 
did not show a clear shell gap at $N=184$ especially for heavier SHE.
This was attributed to the difference in the shell structure 
in the above two approaches. This may have been due to a different 
isospin dependence in the Skyrme model as compared to the RMF theory
\cite{SLK.94,Rein.95}. In comparison, all the Skyrme forces 
have predicted a broad valley of shell stabilization around $Z=120$ and 
$N=$ 172 and 184. On the other hand, Cwiok {\it et al.} \cite{Cwiok.05} 
have suggested a shift in the island of stability 
towards $^{310}_{184}$126.

In spite of an impressive agreement with experimental data for the heaviest
elements, the theoretical uncertainities are large when 
extrapolating to unknown regions of the nuclear chart. 
In particular, there is no consensus among
theories with regard to the regions of the shell stability for superheavy
nuclei. Since in these nuclei the single-particle level density is relatively
large, small shifts in the position of single-particle levels (e.g.,due to the
Coulomb or spin-orbit interaction) can be crucial for determining the shell
stability of a nucleus. While most macroscopic-microscopic
(non-self-consistent) models predict $Z=114$ to be magic, most
self-consistent calculations suggest that the center of the proton shell
stability should be moved up to higher proton numbers, i.e., $Z=120, 124$, or
$126$. For neutrons most non-relativistic calculations predict magic gap
at $N=184$ while the relativistic mean-field theory yields $N=172$ - due to a
slightly different spin-orbit interaction \cite{Lala.96}. 
The experimental determination of superheavy nuclei from this region 
will thus be very important for pinning down the fundamental 
question of the spin-orbit force.

A number of theoretical studies have been carried out within the framework 
of the RMF theory \cite{Gambhir.05,Zhang.05,Sharma.05} to investigate
properties of superheavy nuclei, in particular, $\alpha$-decay properties.
In ref. \cite{Zhang.05} the nature of possible magic numbers within 
the framework of spherical relativistic 
Hartree-Bogoliubov theory using various interactions
has been studied. A recent theoretical work \cite{Bender.01} sheds some 
light on the question of magic superheavy nuclei. According to the
calculations, the patterns of single-particle levels are significantly 
modified in superheavy elements. Firstly, the overall
level density grows with mass number A, as $\alpha$ $A^{1/3}$. Secondly, no
pronounced and uniquely preferred energy gaps appear in the spectrum. This
shows that shell closures which are to be associated with large gaps in the
spectrum are not robust in superheavy nuclei. The theory predicts that
beyond $Z=82$ and $N=126$ the usual localization of shell effects at magic
numbers is diminished. Instead, theory predicts fairly wide areas of
large shell stabilization without magic gaps. 

The Relativistic Mean-Field (RMF) theory \cite{SW.86} has proved to be 
successful as a framework for description of various facets of 
nuclear properties. In the RMF theory, the nuclear force is produced by a
virtual exchange of various mesons. The nuclear saturation is achieved 
by a balance between an attractive $\sigma$- and a repulsive 
$\omega$-field. The relativistic Lorentz covariance of the theory 
allows an intrinsic spin-orbit interaction based upon exchange of
$\sigma$- and $\omega$-mesons. This has been shown to be advantageous 
for properties which depend upon spin-orbit potential.
An immediate advantage of the proper spin-orbit potential has been the
success  of the RMF theory to be able to describe the 
anomalous kink in the isotope shifts of Pb nuclei \cite{SLR.93}.
The isotopic shifts of Pb have been measured with a 
high precision using atomic beam laser spectroscopy and are 
known to show a pronounced kink about the magic
number $N=126$. The non-relativistic approaches based upon 
the Skyrme and Gogny forces have, on the other hand, been unable  
to reproduce this kink. It was shown that this difference in the 
predictions of the Skyrme model and the RMF theory is due
to the isospin dependence of the spin-orbit term \cite{SLK.94,Rein.95}.
The isospin dependence in the RMF theory is provided mainly 
by the coupling constant $g_\rho$ of the $\rho$-meson. 
However, as the strength of the spin-orbit term derives
from a large sum of the absolute values of the scalar and vector fields,
the contribution of $\rho$-field  is much weaker in comparison. Consequently, 
the isospin dependence of the spin-orbit potential in the RMF theory 
is relatively weak \cite{SLK.94}. With the success of the RMF theory in
properties of nuclei related to the shell effects, we have carried
out a study of the superheavy nuclei with an RMF Lagrangian model
which has been shown to have improved shell effects in nuclei.

In the present work, we have explored the region of superheavy nuclei within
the framework of the relativistic mean-field theory using the Lagrangian
model with the vector self-coupling of the $\omega$ meson. 
The region encompasses
the superheavy elements from $Z=102$ up to $Z=120$. Nuclides in these 
isotopic chains with neutron number spanning $N=150$ to $N=190$ have 
been considered. We have used the force NL-SV1 due to Sharma et al. 
\cite{Sharma.00} with the vector self-coupling of the $\omega$ meson. 
Our focus is primarily to look for possible signatures of neutron shell 
gaps or magic numbers in the region of superheavy nuclei.

\section{The Relativistic Mean-Field Theory: Formalism}

In the present work, we have employed the RMF theory to study properties
of superheavy nuclei. The starting point of the RMF theory is the model 
due to Walecka \cite{SW.86}, which describes the
nucleons as Dirac spinors interacting by the exchange of several mesons:
scalar mesons ($\sigma$) couple to the nucleons ($\psi$) through a Yukawa term
$\bar{\psi}\sigma\psi$ and produce strong attraction, isoscalar vector mesons
($\omega$) couple to the conserved nucleon current $\bar{\psi}\gamma_{\mu}\psi$
and cause almost as strong a repulsion. In addition, there are isovector
$\rho$ mesons which couple to the isovector current and photons to produce the
well-known electromagnetic interaction.

We restrict ourselves to the Hartree treatment, where the $A$ independent 
nucleons with the single-particle spinors $\psi_{i}$ (i= 1,..,A) form a 
Slater determinant and move independently in the meson fields. 
The RMF theory starts with the standard Lagrangian density

\begin{eqnarray}\label{eqn:2.1}
 \mathcal{L} &=& \bar{\psi_i}\{i\gamma^{\mu}\partial_{\mu}-M\}\psi_{i} 
\nonumber \\
   & & +\frac{1}{2} \partial^{\mu}\sigma\partial_{\mu}\sigma-U(\sigma)-
g_{\sigma}\bar{\psi}_{i}\sigma\psi_{i} \nonumber \\
 & & -{\frac{1}{4}}\Omega^{\mu\nu}\Omega_{\mu\nu}+
{\frac{1}{2}}m^2_{\omega}\omega^{\mu}\omega_{\mu}-
g_{\omega}\bar{\psi}_{i}\gamma^{\mu}\omega_{\mu}\psi_{i} \nonumber \\
& &-\frac{1}{4}\vec{R}^{\mu\nu}\vec{R}_{\mu\nu}+
\frac{1}{2}m^2_{\rho}\vec{\rho}^{\mu}\vec{\rho}_{\mu}-
g_{\rho}\bar{\psi}_{i}\gamma^{\mu}\vec{\rho}_\mu\vec{\tau}\psi_{i} \nonumber \\
& & -\frac{1}{4}F^{\mu\nu}F_{\mu\nu}-
 e\bar{\psi}_i\gamma^{\mu}A_{\mu}{\frac{(1-\tau_{3})}{2}}\psi_{i}\nonumber   \\
& & +\frac{1}{4} g_4 (\omega_\mu \omega^\mu)^2, 
\end{eqnarray}
where the summation convention is used and the sum over $i$ runs over all
nucleons. Isovector quantities are indicated by the arrow bars. As proposed by
Boguta and Bodmer \cite{Boguta.77}, the $\sigma$ meson moves in a nonlinear
potential of the form:
\begin{eqnarray}\label{eqn:2.2}
U(\sigma)=\frac{1}{2}m_\sigma\sigma^2+\frac{1}{3}g_2\sigma^3
 +\frac{1}{4}g_3\sigma^4.
\end{eqnarray}
Here, $M$, $m_\sigma$, $m_\omega$ and $m_\rho$ are 
the nucleon, the $\sigma$-, the
$\omega$-, and the $\rho$- meson masses, respectively. $g_\sigma$,
$g{_\omega}$,$g{_\rho}$,and $e{^2}$/4$\pi$=1/137 are the coupling constants
for the $\sigma$-,the $\omega$-,the $\rho$-mesons and for the photon. The
coupling constant $g_4$ represents the vector 
self-coupling of $\omega$-meson as introduced in ref. \cite{Sharma.00}.

The classical variation principle provides the Klein-Gordon
equations for the mesons:
\begin{eqnarray}\label{eqn:2.3}
\{-\Delta+m^2_\sigma\}\sigma(\bf{r}) & =&
-g_\sigma\rho_s(\bf{r})- \mathit{g}_2\sigma^2(\bf{r})- 
 \mathit{g}_3\sigma^3(\bf{r}) \nonumber \\
\{-\Delta+m^2_\omega\}\omega^\mu(\bf{r}) &=& g_\omega j^\mu(\bf{r}) +
\mathit{g}_ 4\omega^3 (\bf r) \\
\{-\Delta+m^2_\rho\}\vec\rho^\mu(\bf{r}) &=& g_\rho j^\mu(\bf{r}) \nonumber \\
-\Delta A^\mu(\bf{r})&=& ej^\mu_p(\bf{r}), \nonumber
\end{eqnarray}
whereby the sources are determined by the corresponding density and current
distributions in the static nucleus.  Variation for the 
nucleonic field gives the Dirac equation for nucleons:
\begin{eqnarray}\label{eqn:2.4}
\{-i\alpha.\nabla+\beta M^*(r)+V(r)\}\psi_i(r) &= &\epsilon_i\psi_i(r).
\end{eqnarray}
The effective mass $M^*$(r) is determined by the scalar field $\sigma$(r) as
\begin{eqnarray}\label{eqn:2.5}
M^*(r)&=& M+g_\sigma\sigma(r),
\end{eqnarray} 
and the potential vector potential $V({\rm r})$ is obtained as:
\begin{eqnarray}\label{eqn:2.6}
 V({\bf r})&=& g_\omega\omega_0({\bf r})+g_\rho\tau_3 \rho_0({\bf r})+
e\frac{(1-\tau_3)}{2} A_0({\bf r}).
\end{eqnarray}
Bold faced letters indicate vectors in the 3-dimensional space.
These fields are the solutions of the inhomogeneous Klein-Gordon equations.
As is the case in the majority of applications, the
contributions of antiparticles are neglected, i.e., the {\it no-sea}
approximation is employed. Equation (\ref{eqn:2.3}) together with
Eq. (\ref{eqn:2.4}) provides a closed set of equations.
In order to describe the ground-state properties of nuclei, static 
solutions are obtained from the equations of motion 
Eqs. (\ref{eqn:2.3}-\ref{eqn:2.4}). In this case, the nucleon spinors 
are the eigenvectors of the stationary Dirac equation, which yields 
the single-particle energies $\epsilon_{i}$ as eigenvalues.
The solution of these equations is obtained  iteratively. 
Using these solutions, one calculates physical quantities 
such as the total energy, charge radii
and quadrupole moments amomg others.

\subsection{Axially deformed RMF}

For nuclei in the superheavy region exhibit significant deformations,
we have employed the RMF theory for axially deformed configuration
of nuclei. We solve the Dirac equation (\ref{eqn:2.3}) as well as the
Klein-Gordon equations (\ref{eqn:2.4}) by expansion of the 
wavefunctions into a complete set of eigen solutions of an 
harmonic oscillator potential \cite{GRT.90}. In the axially symmetric 
case the spinors $f^\pm_i$ and $g^\pm_i$ are expanded in terms of the 
eigenfunctions of a deformed axially symmetric oscillator potential 
\begin{equation}\label{eqn:2.47}
V_{osc}(z,r_\bot)=\frac{1}{2}M\omega^2_zz^2+\frac{1}{2}M\omega_\bot^2r_\bot^2.
\end{equation}
Taking volume conservation into account, the two oscillator frequencies
$\hbar\omega_\bot$ and $\hbar\omega_z$ can be expressed in terms of a
deformation parameter $\beta_0$.
\begin{equation}
\hbar\omega_z=\hbar\omega_{0}exp\Bigg(-\sqrt{\frac{5}{(4\pi)}}\beta_0\Bigg)
\end{equation}
\begin{equation}
\hbar\omega_\bot=\hbar\omega_{0}exp\Bigg(+\frac{1}{2}
\sqrt{\frac{5}{(4\pi)}}\beta_0\Bigg).
\end{equation}
The corresponding oscillator length parameters are given by
\begin{equation}
b_z=\sqrt{\frac{\hbar}{M\omega_z}}\qquad {\rm and} \qquad
b_\bot=\sqrt{\frac{\hbar}{M\omega_\bot}}.
\end{equation}
The volume conservation gives $b_\bot^2b_z=b_0^3$. The basis is now determined 
by the two constants $\hbar\omega_0$ and $\beta_0$, which 
are chosen optimally. 

The deformation parameter of the oscillator basis $\beta_0$ is 
chosen to be identical for the Dirac spinors and the meson fields.  
The deformation parameter $\beta_2$ is obtained from the calculated
quadrupole moments for protons and neutrons through
\begin{eqnarray}
Q=Q_n+Q_p&=&\sqrt{\frac{16\pi}{5}}\frac{3}{4\pi}AR_0^2\beta_2
\end{eqnarray}
with $R_0=1.2A^{1/3}$ (fm). The quadrupole 
moments are calculated by using the expression:
\begin{eqnarray}
Q_{n,p}=\langle 2r^2P_2(cos\theta)\rangle_{n,p}
=2\langle 2z^2-x^2-y^2 \rangle_{n,p}.
\end{eqnarray}

\subsection{ Shell effect in nuclei}

Shell effects play an important role in identifying and consequently
synthesizing superheavy elements due to additional stability
provided by the shell effects. Hence few comments on shell effects
would be in order. The shell effects manifest strongly 
in nuclei with the existence of magic numbers. 
The origin of shell closure has long been understood 
due to spin-orbit coupling and to an ensuing splitting of levels.
The spin-orbit potential and a bunching of levels create shell 
closures (magic numbers) which are predicted correctly by most of 
the models. In the non-relativistic density-dependent theory of the 
Skyrme type, the spin-orbit interaction is added phenomenologically and its
strength is adjusted to reproduce the spin-orbit splitting in $^{16}$O. In the
RMF theory, on the other hand, the spin-orbit interaction arises naturally due
to exchange of $\sigma$ and $\omega$ mesons between the nucleons. 
The strength of the  interaction is determined by the spin-orbit 
splitting in $^{16}$O and other nuclei, which is uniquely decided 
by the effective mass. As mentioned in the introduction, the shell effects
in the RMF theory have been advantageous in a few aspects such
as isotope shifts in nuclei. In view of this, application of the
RMF theory to superheavy nuclei may have more realistic predictions.

\subsection{Lagrangian with the vector self-coupling of $\omega$-meson}

In this work, we have employed the Lagrangian model with the quartic
coupling of $\omega$ meson. The non-linear vector self-coupling of 
$\omega$-meson was introduced by Bodmer \cite{Bodmer.91} and 
properties of nuclear matter were discussed on adding a quartic term 
in $\omega$-meson potential. The inclusion of the vector self-coupling 
of $\omega$ meson in addition to the non-linear scalar self-coupling of 
$\sigma$ meson was found to have the effect of softening 
the high-density equation of state (EOS) of the nuclear matter 
\cite{Sugahara.94}. Subsequently, the force NL-SV1 \cite{Sharma.00} 
with the inclusion of the vector self-coupling of $\omega$ meson was 
developed with a view to improve the predictions of the ground-state 
properties of nuclei, such as binding energies, charge radii and 
isotopes shifts of nuclei along the stability line. It was shown that 
with the Lagrangian model NL-SV1, the shell gaps in nuclei along the 
stability line such as Ni and Sn isotopic chains were improved as 
compared to Lagrangian model with the nonlinear scalar coupling alone 
\cite{Sharma.00}. With the improved shell structure along
the line of $\beta$-stability with NL-SV1, it is expected that it shall
have improved predictions in the region far away from the stability line.
In view of this, we have used the parameter set NL-SV1 with the vector 
self-coupling of $\omega$ meson in our investigations. The parameters of the
parameter set NL-SV1 \cite{Sharma.00} are given in Table I.

\begin{table}
\begin{center}
\caption {\label{table:}The Lagrangian parameters of the force NL-SV1 
\cite{Sharma.00} used in the RMF calculations.}
\vglue0.7cm
\begin{tabular}{|l|r|}
\hline\hline
Parameters & NL-SV1~~~~\\
\hline\hline
~~~~$M$ ~~~~~&~~~~~939.0~~~~\\
~~~~$m_\sigma$~~~~~&~~~~~ 510.0349~~~~\\
~~~~$m_\omega$~~~~~&~~~~~ 783.0~~~~\\
~~~~$m_\rho$~~~~~&~~~~~   763.0~~~~\\
~~~~$g_\sigma$~~~~~&~~~~~  10.1248~~~~\\
~~~~$g_\omega$~~~~~&~~~~~ 12.7266~~~~\\
~~~~$g_\rho$~~~~~&~~~~~ 4.4920~~~~\\
~~~~$g_2$~~~~~&~~~~~  $-$9.2406~~~~\\
~~~~$g_3$~~~~~&~~~~~ $-$15.388~~~~\\
~~~~$g_4$~~~~~&~~~~~ 41.0102~~~~\\
\hline\hline
\end{tabular}
\end{center}
\end{table}

\section{Details of calculations}

The input parameters required to carry out explicit numerical calculations
are: neutron pairing gap $\triangle_n$, the proton pairing gap $\triangle_p$
and the number of oscillator shells N$_F$ and N$_B$ of the fermionic
wavefunctions and meson fields, respectively. Both the fermionic and bosonic 
wavefunctions have been expanded in a basis of 20 harmonic oscillator shells 
in this work.

We have carried out a study of superheavy nuclei from No ($Z=102$) to
$Z=120$. The neutron number ranges from $N=150$ to $N=190$. This region
encompasses the neutron number $N=184$ which has been predicted to be a 
magic number in various theories. A possible deformed shell has also been 
predicted in this region as discussed in Section I. We have carried out 
deformed RMF+BCS calculations for superheavy nuclei in the above region. 
For pairing gaps, we have used the formula due to 
M\"oller and Nix \cite{MN.92} as given by:
\begin{equation}
\triangle_ {n} = 4.8 ~N^{-1/3}
\end{equation}
\begin{equation}
\triangle_{p} = 4.8  ~Z^{-1/3}.
\end{equation}
An axially symmetric deformed configuration with reflection symmetry has been
assumed for nuclei. For each nucleus an initial basis deformation of prolate
as well as oblate type has been taken in order to seek a minimum both in
the prolate as well as in the oblate region of the deformation space. 
Consequently, we have obtained a prolate and an oblate minimum for most of 
the nuclei. However, nuclei close to a potential magic number assume a 
spherical shape.

\section{Results and Discussions}

\subsection{Binding energies and deformations}

\begin{figure}
\begin{center}
\includegraphics[width=1.0\textwidth,height=0.72\textheight]{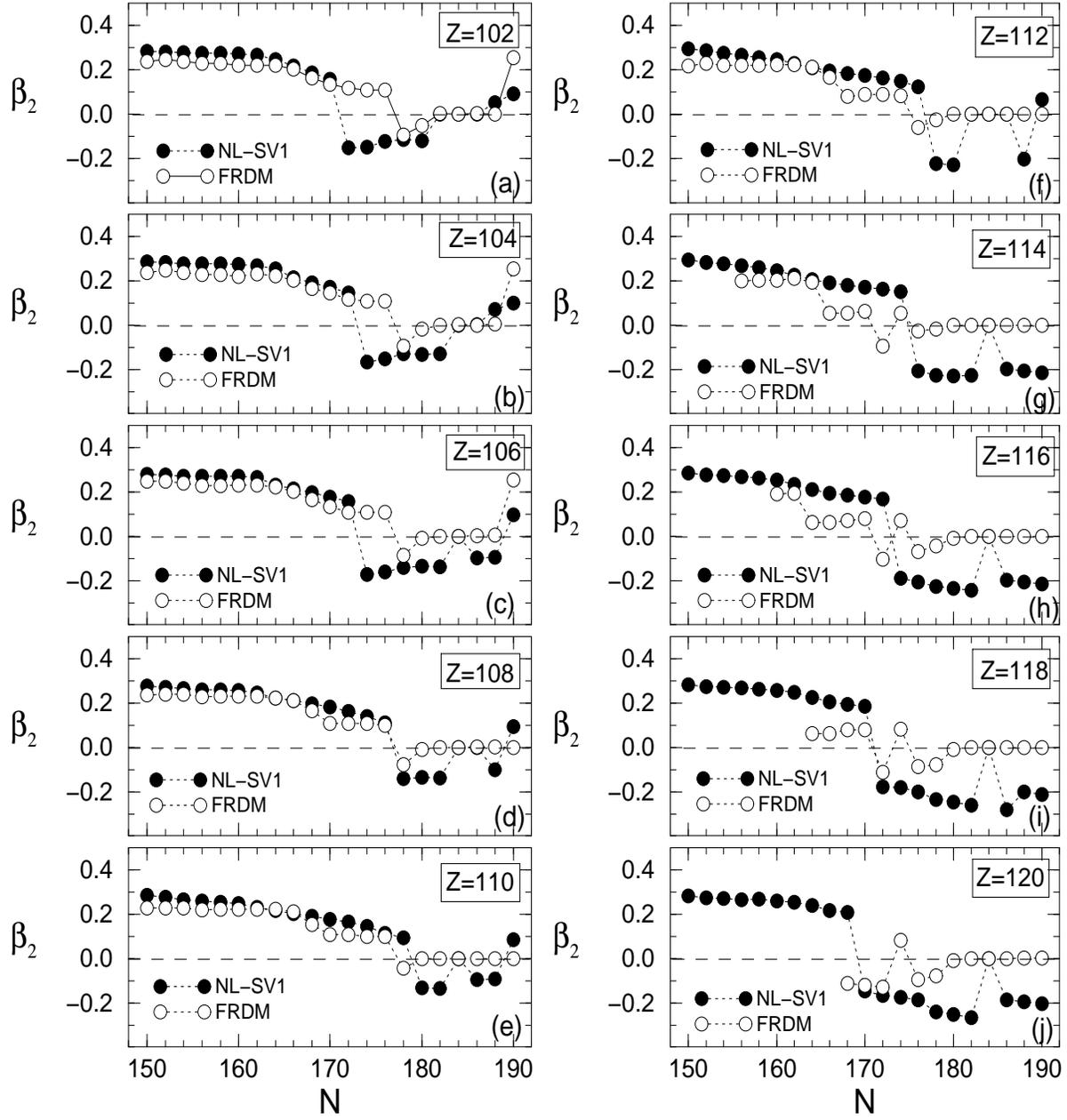}
\caption{The $\beta_2$ values for isotopes of the elements $Z=102-120$ 
obtained with NL-SV1. The results with FRDM are shown for comparison.}
\label{fig:fig1}
\end{center}
\end{figure}

The RMF+BCS minimizations with NL-SV1 give rise to a prolate and an 
oblate solution for most of the nuclei. We show the quadrupole deformation
$\beta_2$ for the lowest energy state (ground state) for all the 
isotopic chains in Figures ~\ref{fig:fig1}(a)-~\ref{fig:fig1}(j).  
A large number of nuclei in these chains assume a prolate shape in the 
ground state. This is especially the case in the region of 
neutron number $N\sim150-170$. Typically, these nuclei exhibit a 
prolate deformation of $\beta_2 \sim 0.2-0.3$. 

For the isotopic chains of No $(Z=102)$, Rf $(Z=104)$ and Sg $(Z=106)$, 
there is a transition from a prolate shape to an oblate shape 
at $N\sim 170$. Beyond $N=172$, nuclei assume an oblate shape in 
the ground state for these isotopic chains (see
Figs.~\ref{fig:fig1}(a)$-$~\ref{fig:fig1}(d)). 
The magnitude of the oblate deformation decreases as the neutron number 
increases above $N=172$, which approaches a vanishing value as the 
neutron number $N=184$ arrives. In the vicinity of $N=184$, nuclei
exhibit a spherical shape.  Such a feature has been commonly predicted 
by several relativistic as well as non-relativistic theories, 
whereby $N=184$ constitutes a major magic number in the region of
superheavy nuclei.

As one goes to the heavier elements above Sg $(Z=106)$, one can 
visualize that the tendency of prolate shapes in the range 
of neutron numbers $N=150-176$ continues for heavier superheavy 
elements as well. The point of shape transition from 
prolate to oblate seems to be shifting to a higher neutron
number. For instance, for Hs ($Z=108$) the prolate to oblate
transition occurs at $N=178$.  For the isotopic chain Ds ($Z=110$)
the shape transition point advances to $N=180$.
However, for $Z=112$ and above, the transition point falls back 
to $N=178$ and below it. For instance, for $Z=116$ and $Z=118$, 
the shape transition occurs at $N=174$ and $N=172$, respectively. 
Similarly, the shape transition point for $Z=120$ falls back to 
$N=170$. Thus, the number of nuclei which exhibit an oblate shape 
below the neutron number $N=184$ increases steadily in going from
$Z=112$ to $Z=120$. As far as the spherical shape for $N=184$ nuclei 
is concerned, it is exhibited by all the isotopic chains from 
$Z=102$ to $Z=120$. This is reminiscent of a magic shell character 
for the neutron number $184$.

The general feature of shape transition from prolate to oblate in the region of
$N=172-176$ prevails for all the isotopic chains considered in this study.
We have made a comparison of NL-SV1 deformations with those predicted by the
macroscopic-microscopic mass formula FRDM. For the isotopic chains 
$Z=102-112$, the FRDM exhibits a predominantly prolate shape for the 
neutron numbers $N=150-176$. This feature is very similar to that
predicted by NL-SV1. 

The FRDM shows a shape transition from prolate to an oblate shape at the 
neutron number $N \sim 178$ for the chains $Z=102-110$. There is only 
a single nuclide exhibiting an oblate shape in FRDM at $N \sim 178$ for
these chains, beyond which all the nuclei assume a spherical shape 
for the isotopic chains of $Z=102-112$ in the FRDM predictions, in contrast 
to NL-SV1. The oblate deformation for $N=178$ nuclei in this region
amounts to $\beta_2 \sim 0.1$ in the FRDM. This is unlike NL-SV1, where a 
significant number of nuclei exhibit a moderately deformed 
oblate shape until the supposedly magic number $N=184$ arrived at. 

For the isotopic chains $Z=114-118$, the behaviour of FRDM for deformation
properties is rather uneven. For the region of the lower neutron numbers, 
some nuclei in these chains show a moderately prolate shape, 
whereas a few others exhibit a small prolate deformation. 
This is followed by a shape transition to an oblate deformation 
at $N=172$, followed by a spherical shape in going to higher neutron
numbers. That nuclei in the vicinity of $N=184$ are predominantly spherical
in FRDM points to the hypothesis that $N=184$ maintains its magicity 
even for heavier superheavy elements in FRDM.

\subsection{Comparison with ETF-SI results} 

\begin{figure}
\begin{center}
\includegraphics[width=1.0\textwidth,height=0.72\textheight]{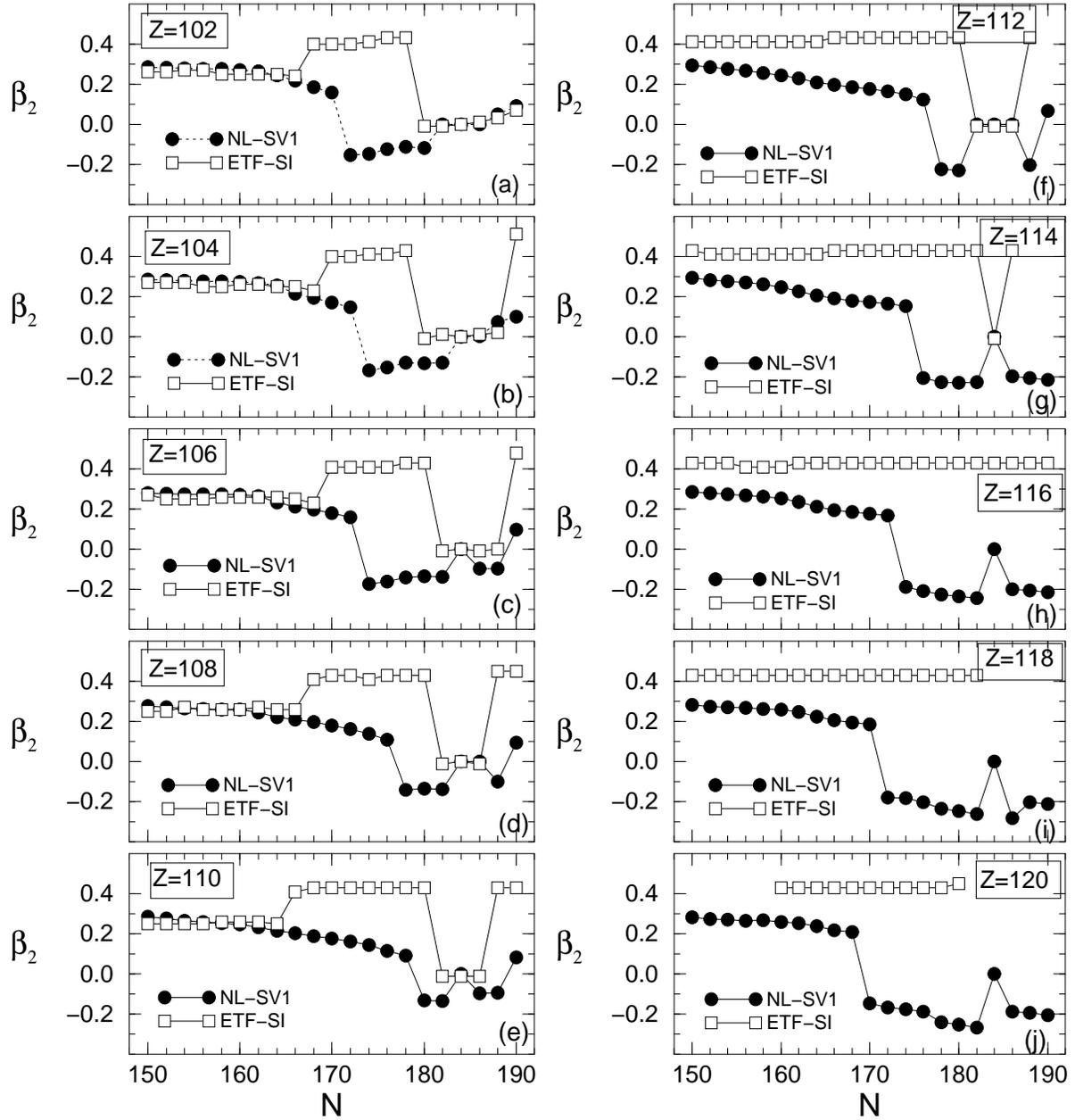}
\caption{The $\beta_2$ values for the isotopes of the elements 
$Z=102-120$ obtained with NL-SV1 are compared with ETF-SI predictions.}
\label{fig:fig2}
\end{center}
\end{figure}

The mass formula Extended Thomas-Fermi with Strutinsky integral (ETF-SI) 
\cite{ETFSI} was obtained to fit experimental masses of more 
than 1600 nuclei from all over the periodic table. 
The ETF-SI, due to its partly microscopic structure, has found
broader acceptance for nuclear physics applications. It predicts binding
energies and deformations of nuclei over a broad range of masses quite well.
Here we compare the ground state deformations obtained with NL-SV1 with those
predicted by ETF-SI. The quadrupole deformations $\beta_2$ are compared for
the given isotopic chains in Figs.~\ref{fig:fig2}(a)$-$\ref{fig:fig2}(j).
Figs.~\ref{fig:fig2}(a)$-$\ref{fig:fig2}(d) show that ETF-SI deformations
are in good agreement with the NL-SV1 values from $N=150$ to 
$N \sim 166$ for the elements from Ds ($Z=102$) to Hs ($Z=108$). 
For Ds ($Z=110$), the agreement between NL-SV1 and ETF-SI is 
good up to $N=164$ (see Fig.~\ref{fig:fig2}(e)). 

A constrasting picture with the ETF-SI is that for neutron numbers 
$N \sim 166$ and above, it shows a strong divergence from the NL-SV1 
predictions. ETF-SI does not exhibit a prolate to oblate transition
as do NL-SV1 and FRDM predictions. In constrast, ETF-SI predicts a 
constant and a large value of $\beta_2 \sim 0.4$ for nuclei in 
the range of $N \sim 168-180$ for the isotopic chains of $Z=102-110$. 
For the heavier elements $Z=112-120$, ETF-SI predicts a nearly
constant value $\beta_2 \sim 0.4$ for all the neutron numbers $N=150-180$.

ETF-SI predicts a spherical shape of nuclei in the vicinity of $N=184$ 
for the elements $Z=102-114$. This feature of ETF-SI is similar to 
that of FRDM. However, for $Z=116$, ETF-SI bypasses the spherical
shape for $N=184$. On the other hand, for several isotopic chains ETF-SI
predicts a shape transition from spherical to a strongly prolate shape 
at $N=188-190$ (see Figs.\ref{fig:fig2}(b)$-$~\ref{fig:fig2}(g)).

For $Z=110$ and $Z=112$, nuclei with two neutrons below and two neutrons 
above $N=184$ exhibit a spherical shape, whereas for $Z=114$ 
there is only one nuclide i.e,
$N=184$ that shows a spherical shape. It's neighbours on both sides,
assume a prolate shape in ETF-SI. This is indicative of a decrease in the shell
strength at $N=184$ with ETF-SI. This point is supported by a deformed 
shape taken by the $N=184$ nuclide for $Z=116$ (See Fig.~\ref{fig:fig2}(h).)
with ETF-SI. Whilst ETF-SI values for $Z=118$ and $Z=120$ are not known, 
the aforesaid diminishing of $N=184$ shell strength points to an 
erosion of a possible $N=184$ magicity for the heaviest superheavy 
elements from $Z=116$ to $Z=120$ with ETF-SI.

\begin{figure}
\vspace{0.5cm}
\begin{center}
\includegraphics[width=0.90\textwidth]{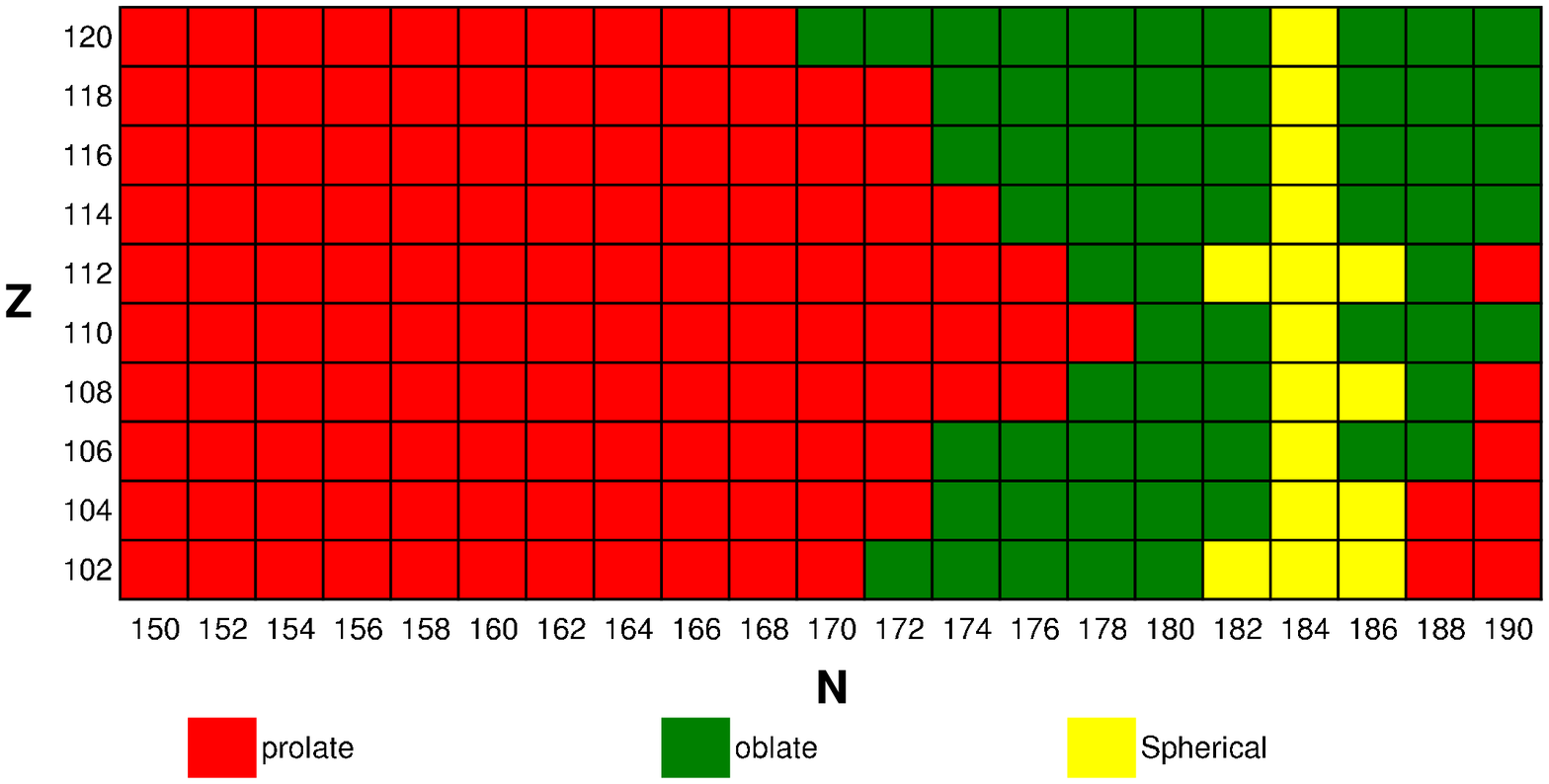}
\caption {The landscape of the quadrupole deformation $\beta_2$
 obtained with NL-SV1 for $Z=102-120$ and $N=150-190$.}
\label{fig:fig3}
\end{center}
\end{figure}
A summary of quadrupole deformation $\beta_2$ in the ground state of superheavy
nuclei with NL-SV1 obtained in this work is shown in the Fig~\ref{fig:fig3}. 
The nuclei denoted by the red colour indicate a prolate ground state. 
An oblate ground state is shown by the green colour. Spherical nuclei are
indicated by yellow. As seen in Fig.~\ref{fig:fig3}, 
nuclides with $N=150-172$ for almost all the isotopic chains exhibit a 
prolate shape. On the other hand, nuclides with $N > 172$ exhibit an 
oblate shape until a spherical shape appears for nuclei close to $N=184$. 
There is a little exception here in that for SHE with
$Z=108$ to $Z=112$, a few nuclides above $N=172$ continue to exhibit a prolate
shape. This indicates an intrusion of prolate shape into the domain of oblate
shapes as indicated in the middle of the figure.
Nuclei in the vicinity of $N=184$ are predominantly spherical as shown by the
yellow blocks. In a few cases and especially for $Z=102$, nuclei two neutrons
below and two neutrons above $N=184$ assume a spherical shape. As shown by the
picture, $N=184$ acts as a major magic number in the neutron-rich region 
for all the SHE with $Z=102-120$. 

As we will see in the single-particle levels, the neutron
number $N=184$ remains as a major magic number. Also we will see that 
in the RMF theory with NL-SV1, the proton number $Z=120$ is also a 
magic number. The combination $Z=120$ and $N=184$ seems to act as 
double closed shell in the RMF theory.

Nuclides above $N=184$ show a prolate (red) region for a few lighter
SHE. However, for heavy SHE with $Z=108$, nuclides above $N=184$ are 
predominantly oblate (green) as shown in the Fig.~\ref{fig:fig3}. 
Thus, a significant region below as well as above $N=184$ lends itself 
to an oblate deformation.

\subsection{Shape co-existence in the superheavy region}

The phenomenon of shape co-existence in the region of deformed nuclei
arises due to interplay of deformed single-particle levels. Splittings and
alignment of deformed single-particle levels leads often to ground states
with a prolate and an oblate shape nearly degenerate in energy. For practical
purposes, when the two minimum energy states are within $\sim$1 MeV, the two
states are said to be shape co-existent.
\begin{figure}
\begin{center}
\includegraphics[width=0.95\textwidth]{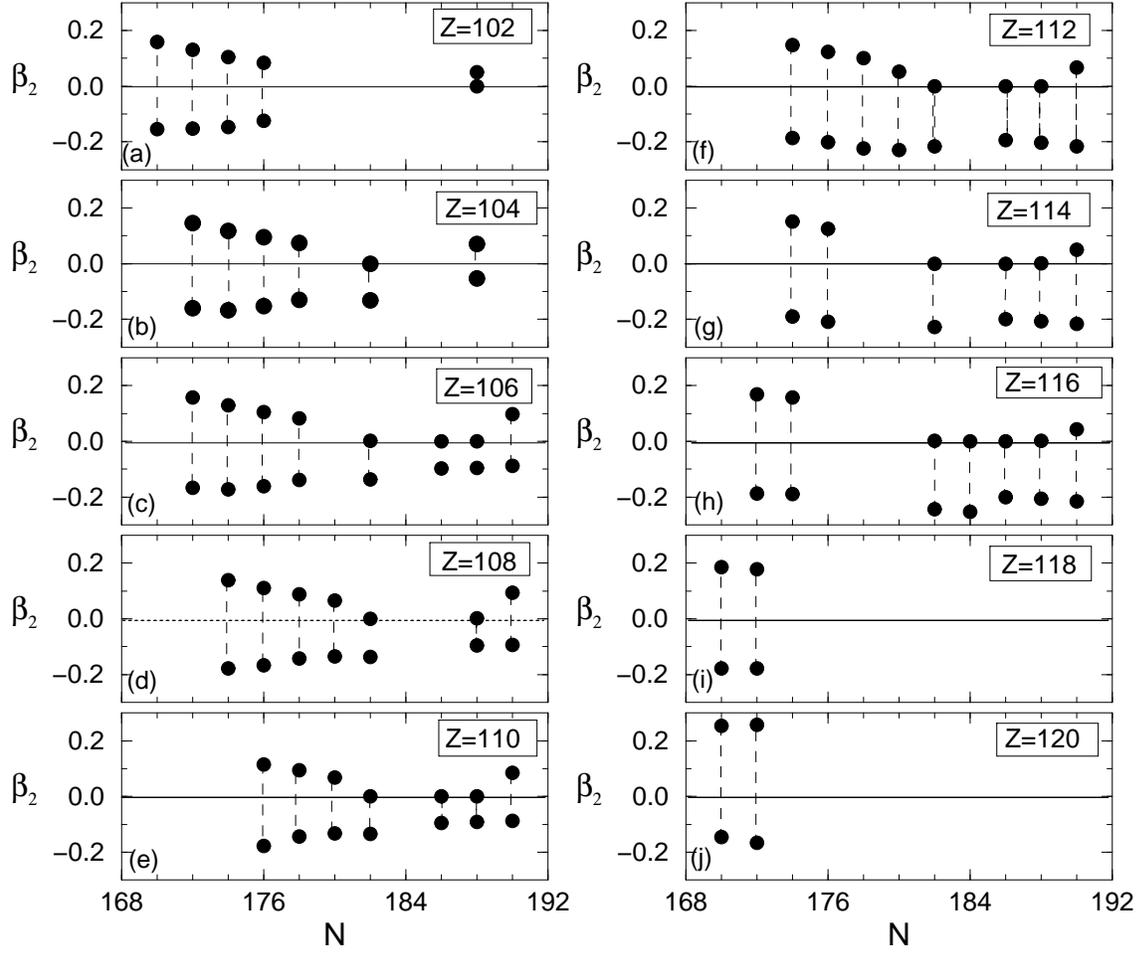}
\caption{Isotopes of $Z=102-120$ exhibiting the shape-coexistence 
in the Lagrangian model NL-SV1.}
\label{fig:fig4}
\end{center}
\end{figure}

We show in Figs.~\ref{fig:fig4}(a)$-$\ref{fig:fig4}(j) nuclides 
which exhibit a shape co-existence with NL-SV1.
Looking at the figures one can see that a large number of nuclides 
exhibit the shape co-existence for the superheavy chains. 
The nuclides in isotopic chains from $Z=102$ (No) to $Z=112$ 
show a common pattern of shape co-existence between prolate and oblate 
shapes for the region of neutron numbers $N \sim 170$ to $N \sim 182$. 
Whilst for $Z=102$ and $Z=104$ shape co-existence is limited to 
neutron numbers $N \sim 168$ to $N \sim 178$, for $Z=106$ to $Z=112$ 
the region of shape co-existence extends to  neutron number $N=182$. 
Whereas for the lighter neutron numbers there is a prolate-oblate 
shape co-existence for elements $Z=102$ to $Z=112$, there is a
shape co-existence of spherical-oblate shapes for $N=182$ isotones from
$Z=104-112$. This indicates a softer nature of $N=184$ shell closure even 
for the lighter superheavy elements. Several nuclides with $N > 184$, 
also exhibit a spherical-oblate shape co-existence for the isotopic 
chains $Z=106-112$. 

Figures ~\ref{fig:fig4}(g)-~\ref{fig:fig4}(j) show the shape
co-existence for heavier superheavy elements from $Z=114$ to $Z=120$.
Unlike the picture for superheavy elements $Z=102-112$, the shape
co-existence in the lighter neutron region is reduced significantly for
$Z=114-120$. Only a few nuclides below $N=180$ exhibit a prolate-oblate
shape co-existence. For $Z=114$ and $Z=116$, several nuclides exhibit a
spherical-oblate shape co-existence for neutron numbers $N>184$. 
Interestingly, $N =184$ nuclide for $Z=116$ shows a spherical-oblate shape
co-existence. This indicates a softening of the shell effects 
for $N=184$ for $Z=116$. 

The superheavy elements $Z=118$ and $Z=120$ exhibit the shape co-existence 
rather scarcely i.e, only two isotopes $N=170$ and $N=172$ show a 
prolate-oblate shape co-existence. In the rest of the neutron region, 
there is no shape co-existence for the heaviest superheavy elements 
with $Z=118$ and $Z=120$ This may imply that the shell strength at
$N=184$ has picked up for these chains and that a readjustment of the
deformed single-particle levels does not lend to degenerate shapes
in the ground state.
\begin{figure}
\vspace{0.5cm}
\begin{center}
\includegraphics[width=0.6\textwidth]{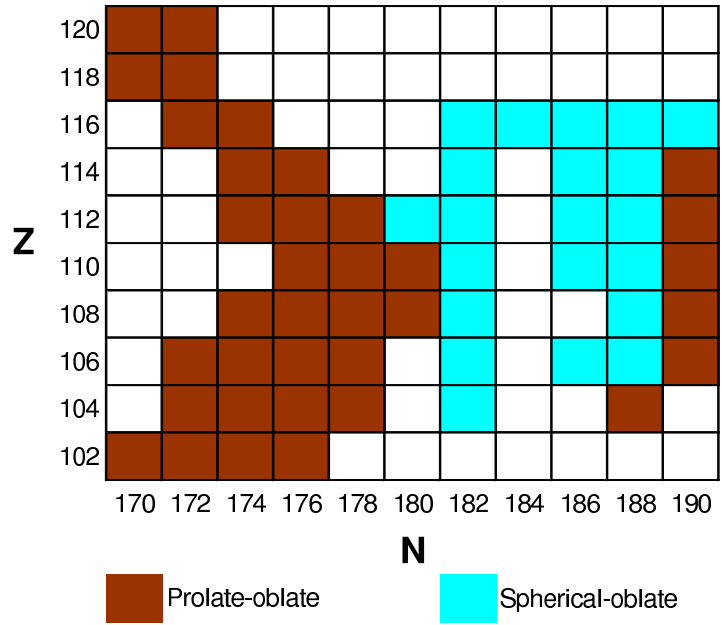}
\caption{The landscape of shape co-existence in superheavy nuclei with NL-SV1.
The prolate-oblate and spherical-oblate shape co-existenec is depicted.}
\label{fig:fig5}
\end{center}
\end{figure}

A summary of the shape co-existence in superheavy nuclei is illustrated in
Fig.~\ref{fig:fig5}. The brown squares depict a shape co-existence between a
prolate and an oblate shape, whereas the blue squares represent a shape
co-existence between a spherical and an oblate shape. As the figure shows,
the phenomenon of prolate-oblate shape co-existence pervades 
the region from $N=170$ to $N=180$. For the lighter elements this shape 
co-existence lies in the neutron region 170-178. The lower boundary 
of this region gradually shifts to higher neutron numbers $Z=110$
is reached. As one moves above $Z=110$, the region shifts 
gradually towards lower neutron numbers again. From $Z=114$ to $Z=120$,
however, there are fewer cases of the prolate-oblate shape
co-existence. Apparently, the region of prolate-oblate shape 
co-existence as shown by the brown colour is symmetric about $Z=110$. 
In addition, there are a few more cases of prolate-oblate 
shape co-existence for higher neutron numbers near N$=190$.

The region of spherical-oblate shape co-existence lies near $N=184$. The
neutron number $N=184$ being a major magic number, the ground-state of
nuclides with $N=184$ is  spherical as seen earlier. 
For almost all the isotopic chains there is no shape 
co-existence (empty squares) at $N=184$, with the 
exception of $Z=116$. Only for $Z=116$, there is a spherical-oblate shape
co-existence at $N=184$. This is due to an erosion of the magicity
of $N=184$ for $Z=116$.

Nuclides with $N=182$ show a spherical-oblate shape co-existence as shown
by blue squares. Nuclides with $N=186$ and $N=188$ also exhibit a
spherical-oblate shape co-existence for several isotopic chains. The fact that
nuclides just below and above $N=184$ display a spherical shape co-existence
indicates that $N=184$ is not such a strong magic number. This can be 
constrasted with other magic numbers in the periodic table which exhibit 
robustness. There are rarely occasions of a shape-coexistence or even
a deformation in the direct vicinity of a major magic number especially
when one is not very far from the line of $\beta$-stability. Superheavy
nuclei especially those associated to $N=184$ do not fall in this category
and are deemed to be in a region far from the stability line. Therefore,
mellowing of the shell strength of $N=184$ is not  unexpected.

\subsection{Two-neutron separation energies }
\vglue0.1cm
Magic numbers in nuclei are characterized by a large shell gap between the
last single-particle level of the magic number and the next single-particle
level above it. This results in a larger value of 2-neutron separation energy
$S_{2n}$ for a nucleus with 2-neutrons more than the magic number. 
Consequently, the major magic numbers exhibit a characteristic kink 
in $S_{2n}$ values all over the periodic table. Thus, the difference 
$S_{2n}(N) - S_{2n}(N+2)$ at the magic neutron number $N$ reflects 
its shell gap. Its magnitude is a measure of the shell strength.
\begin{figure}
\begin{center}
\includegraphics[width=0.62\textwidth]{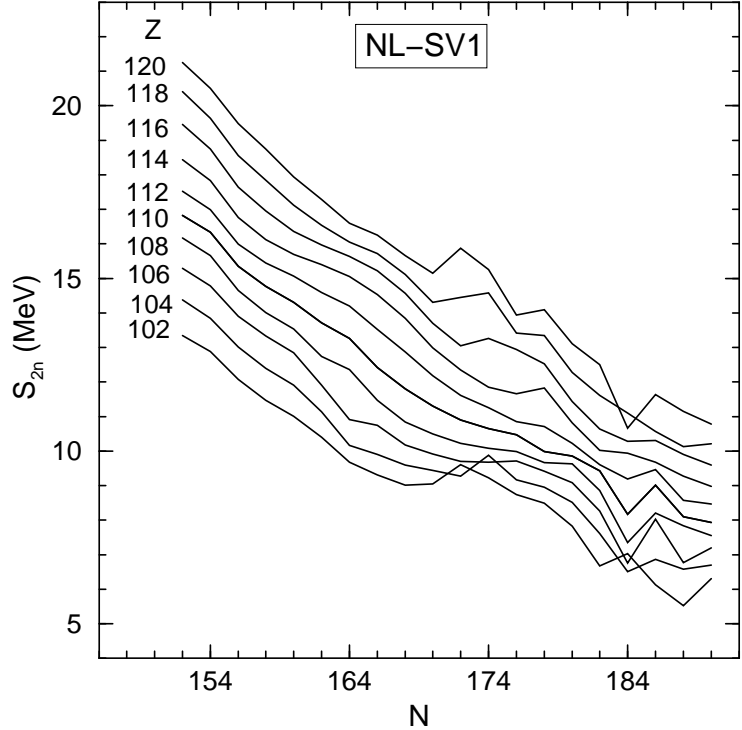}
\caption{Two-neutron separation energy $S_{2n}$ for the isotopes of
$Z=102$ to $Z=120$ with neutron numbers ranging from $N=150$ to  $N=190$.}
\label{fig:fig6}
\end{center}
\end{figure}

With a view to visualize as to whether there are possible shell closures in
neutron numbers for superheavy nuclei, we have computed $S_{2n}$ values for
all the isotopic chains. Using the binding energies of neighbouring isotopes,
S$_{2n}$ is defined as,
\begin{equation}
S_{2n}= B.E.~(N) - B.E~(N+2).
\end{equation}
The $S_{2n}$ values for the isotopic chains considered in this work are shown
in Fig.~\ref{fig:fig6}. The values correspond to the lowest energy state
(ground state)  obtained with RMF+BCS minimization using the force NL-SV1. As
expected, the $S_{2n}$ values show a decreasing trend in going to nuclei with
higher neutrons. The $S_{2n}$ values show a rather monotonous decrease in
going from $N=152$ to $N\sim170$ with a slight kink-like structure in the
neighbourhood of $N\sim154$ and $N\sim164$. Due to significant changes in
deformation properties of nuclei near $N\sim172$ and $N\sim184$, there is
much more structure in the $S_{2n}$ values.
\begin{figure}
\begin{center}
\includegraphics[width=0.80\textwidth]{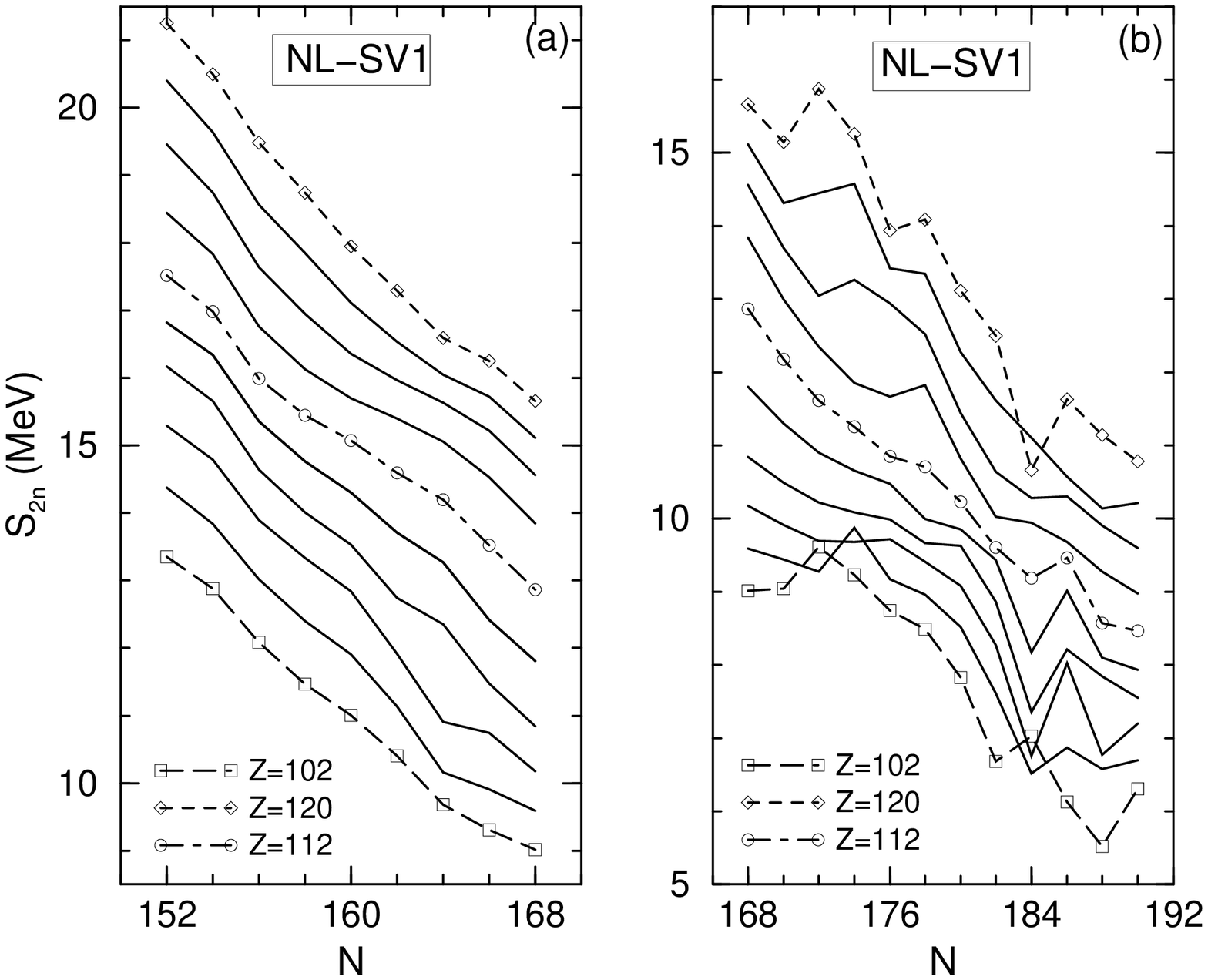}
\caption{(a) Two-neutron separation energy $S_{2n}$ for the isotopes 
of elements $Z=102$ to $Z=120$, with neutron numbers ranging 
from $N=152$ to $N=170$. (b) S$_{2n}$ values for $N=170$ to $N=190$.}
\label{fig:fig7}
\end{center}
\end{figure}

In order to visualize the structure in the $S_{2n}$ values, we have split
Fig.~\ref{fig:fig16} into two parts which are shown in Fig.~\ref{fig:fig7}. In
Fig.~\ref{fig:fig17} (a) the $S_{2n}$ values for $N=152-168$ are shown. 
The higher neutron part with $N=168-190$ is shown  in Fig.~\ref{fig:fig7}
(b). One can notice a slight kink at $N=154$ for several isotopic 
chains. A kink like structure is visible for $Z=106-116$. This structure
diminishes for $Z=118$ and $Z=120$. The structure at $N=154$ is compatible
with a deformed shell gap predicted by non-relativistic 
macroscopic-microscopic approaches \cite{Cwiok.96}.

One can also notice a slight kink at $N\sim164$. Thus, there seems
to exist a few pockets of presumably deformed shell gaps at $N\sim154$ and
$N\sim164$. However, there are no signatures of any strong magic number 
in the region $N=152-168$.  Fig.~\ref{fig:fig7} (b) shows
structures in $S_{2n}$ values at $N \sim$ 172, 178 and 184. The
behaviour of $S_{2n}$ values in the neutron region $N=168-190$ is far from
monotonous. However, for the lighter superheavy elements, there 
is very little structure at $N \sim 172$ except for $Z=102$ and 104. 
For the heavier elements with $Z=116-120$, the kink at 
$N\sim172$ shows a slightly upward trend. A kink like
structure in this region indicates a possible $N\sim172$ deformed shell gap
for superheavy elements. A combination of $N \sim 172$ and $Z \sim 120$
might lend itself to a structure akin to a double shell closure.

The region near $N \sim 184$ is full of structural effects as can be seen in
Fig.~\ref{fig:fig7} (b). A kink in the $S_{2n}$ values can be seen clearly at
$N=184$ in several isotopic chains with $Z=104-110$. Thus, for the lighter
superheavy elements NL-SV1 predicts a shell closure at $N=184$ as indicated
by the kink. For the deformed RMF+BCS calculations with NL-SV1, a sharp 
decrease in the S$_{2n}$ values at $N=184$ is indicative of a shell 
closure which is consistent with that with NL-SH \cite{Lala.96}. For SHE above
$Z \geq $112 the kink and the structure at $N=184$ is reduced significantly. 
Thus, for the heaviest SHE, $S_{2n}$ values do not give an unambiguous 
indication of a shell closure at $N=184$ . 

\subsection{Alpha-decay properties - $Q_\alpha$ values}

In the pursuit of superheavy heavy elements, it is one of aims of research
to find out regions of extra-stability 
with a view to synthesize superheavy elements with
the most appropriate projectile and target combinations. With this in mind, 
it will be helpful if one can explore the regions of higher stability
theoretically. Depending upon the region of an extra-stability which would
arise from shell gaps and possible shell closures, the half-life of
$\alpha$-decay is a potential indicator of a possible value of stability for 
an area of enhanced stability. It is expected that for nuclei near a shell 
closure the $\alpha$-decay half-lives would be larger than their neighbours.

\begin{figure}
\begin{center}
\includegraphics[width=0.60\textwidth]{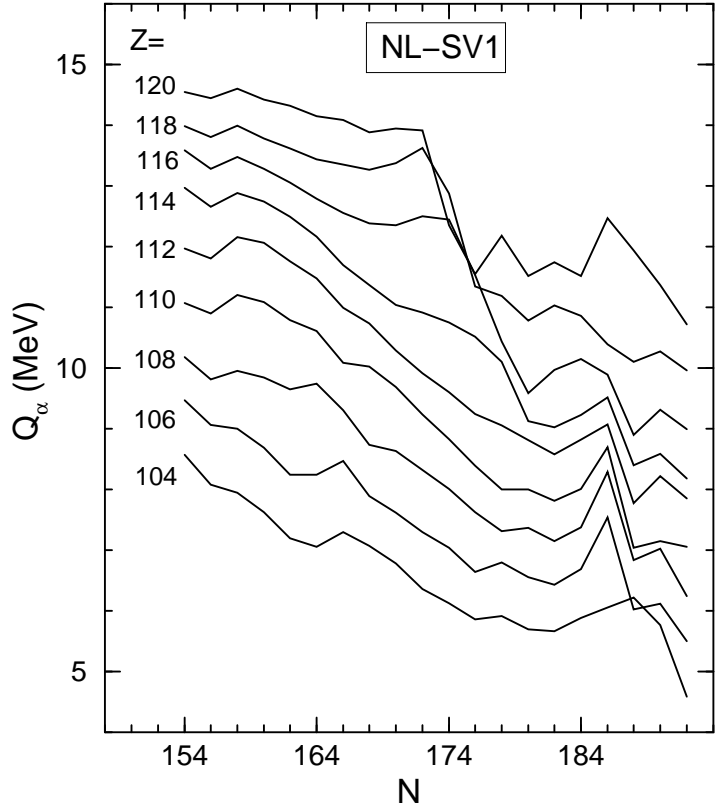}
\caption[$Q_\alpha$ values obtained with NL-SV1]
{Q$_\alpha$ values for the isotopic chains
  with $Z=102-120$ and $N=150-190$ obtained with NL-SV1.}
\label{fig:fig8}
\end{center}
\end{figure}

Nuclei in the transuranium and superheavy region are known to be $\alpha$
emitters. The rate of $\alpha$-decay depends strongly on $Q_{\alpha}$ value of
a nucleus. We have calculated $Q_{\alpha}$ values 
for nuclei in all the isotopic
chains we have considered. The Q$_\alpha$ values thus obtained are 
shown in Fig.~\ref{fig:fig8}. As one moves from $Z=104$ to 
$Z=120$, $Q_\alpha$ values show a continuous increase with increase in 
the charge number of a nucleus. For all the isotopic chains one can see a
decrease in the $Q_\alpha$ values with an increase in neutron number. 
A decreasing $\alpha$-value implies a higher $\alpha$-decay
half-life. Consequently, in moving from the lower neutron number 
$N=150$ towards $N=184$, one has a stabilizing factor with a 
decrease in the $Q_\alpha$ values. For nuclides with $N=186$, 
a slight peak in $Q_\alpha$-values for several isotopic chains 
with $Z=104-114$ can be seen. In terms of $\alpha$-decay half-life 
(as we will see below) it will mean that nuclides with $N=186$ 
become more unstable towards $\alpha$-decay.

In going above $N=184$ one sees a little peak at $N=186$ for 
the isotopic chains with $Z=106-114$, though for the chain with 
$Z=104$, $Q_\alpha$ shows an increasing trend from $N=184$ to $N=186$. 
In terms of $\alpha$-decay half-life this would imply a decreased 
half-life for nuclides with $N=186$. For the isotopic chains 
above $Z=114$ such a behaviour is not seen, but for only $Z=120$. Above 
$N=186$, there is a decrease in $Q_\alpha$ values for all the isotopic chains.

\subsection{Comparison with the experimental data (systematics)}

A large amount of experimental effort to the synthesis of 
superheavy elements as well as to explore the nuclear properties 
of superheavy nuclei by the Lawrence Berkeley National Laboratory, 
Berkeley, GSI, Darmstadt and the Joint Institute for Nuclear 
Research, Dubna. Binding energies (masses), $Q_\alpha$ values and
$\alpha$-decay half-lives of many nuclei have been determined or
estimated. Naturally, such data is accessible only for 
light mass SHE especially with $Z=100-110$. Only few data 
exist for $Z=112$ and $Z=114$. 

\begin{figure}[h]
\begin{center}
\includegraphics[width=0.85\textwidth]{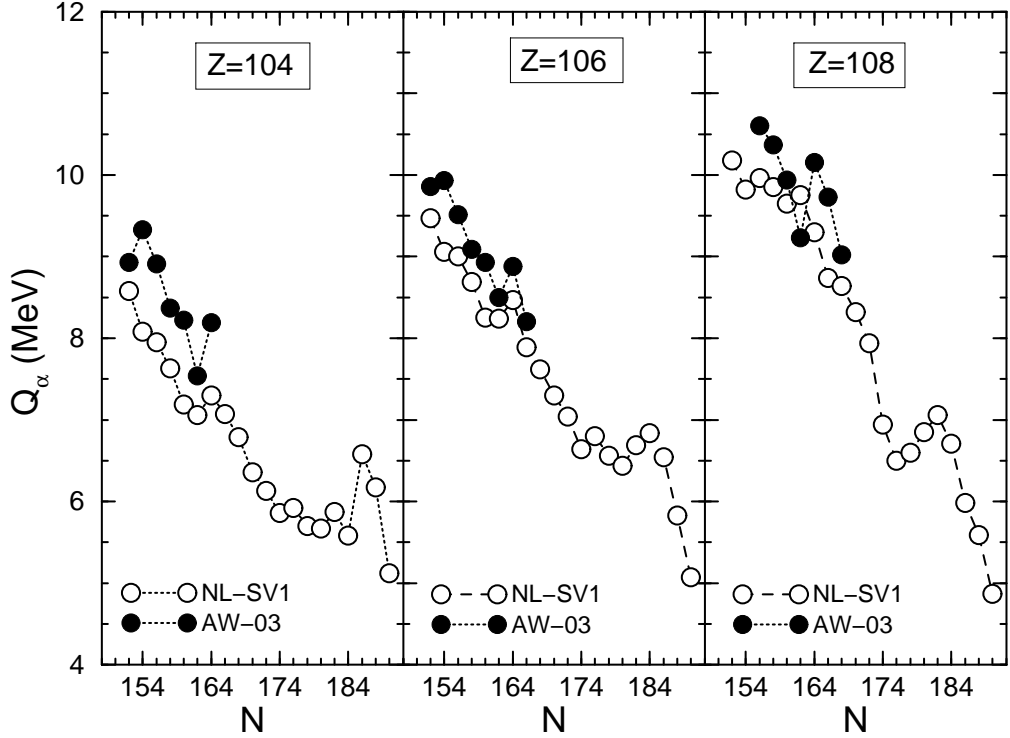}
\caption{Q$_\alpha$ values obtained with NL-SV1
for the isotopic chains  with $Z=104,106$ and 108 
for the neutron numbers ranging from $N=150-190$. 
The experimental values taken from the systematics of 
Audi-Wapstra \cite{Audi.03} are shown for comparison. }
\label{fig:fig9}
\end{center}
\end{figure}
\begin{figure}[h]
\begin{center}
\includegraphics[width=0.85\textwidth]{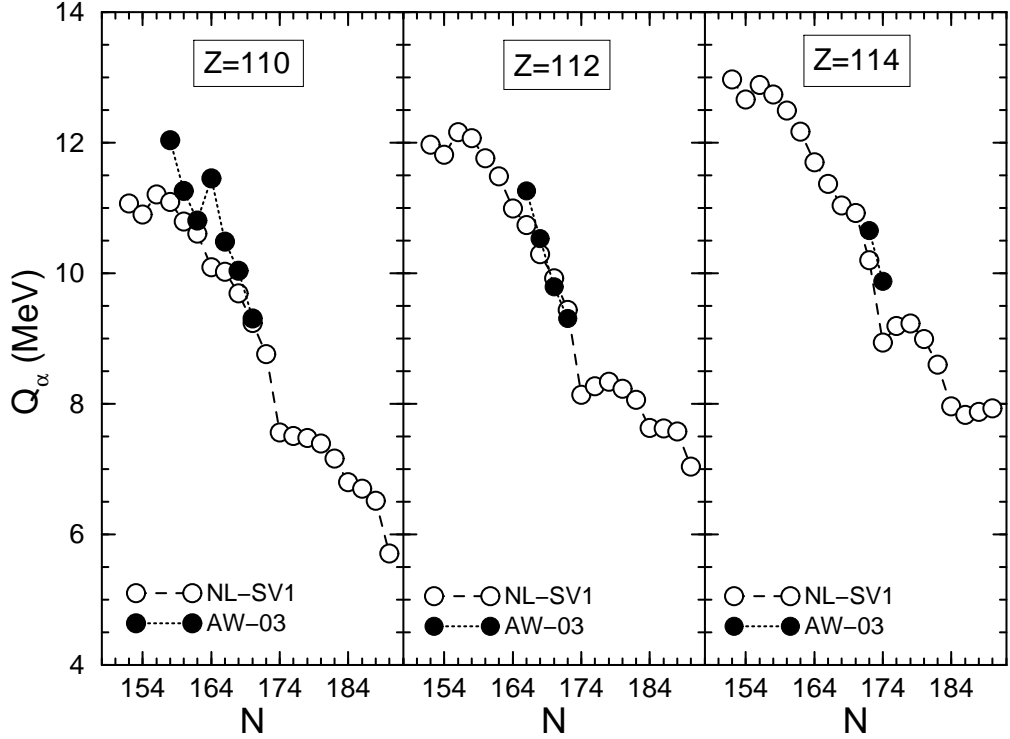}
\caption{$Q_\alpha$ values obtained with NL-SV1
for the isotopic chains of $Z=110,112$ and 114 
for the neutron numbers $N=150-190$. The experimental values 
taken from the systematics compiled by Audi-Wapstra \cite{Audi.03} 
are shown for comparison.}
\label{fig:fig10}
\end{center}
\end{figure}

We show in Figs.~\ref{fig:fig9} and ~\ref{fig:fig10}, $Q_\alpha$ values 
obtained with NL-SV1. These are compared with $Q_\alpha$ values taken 
from 2003 compilation of systematics by Audi-Wapstra \cite{Audi.03}. 
As seen in Fig.~\ref{fig:fig9}, NL-SV1 values underestimate the empirical 
$Q_\alpha$ values for $Z=104-108$ slightly. The disagreement was a bit 
more pronounced for $Z=104$. For $Z=106$ the agreement of 
NL-SV1 values with the data is reasonably good. For the
case of $Z=104$ and $Z=106$, the kink in $Q_\alpha$ 
values at $N=162$ in the experimental data is reproduced 
well by NL-SV1. For the chain of $Z=108$, though NL-SV1 values are
close to the experimental data, the kink at $N=162$ in the experimental data
is not reproduced  by NL-SV1. This may be due  to complex shell 
structure around $Z=108$ than that presented by the mean field in the RMF.

In Fig.~\ref{fig:fig10} the NL-SV1 values show agreement with several
data points, whereas there is a discrepancy of more 
than 1 MeV for the data points of $N=158$ and $N=166$. Such differences 
may arise from a different evolution of shape in reality 
than predicted by the theory. In contrast, the NL-SV1 values show a 
very good agreement with experimental data for SHE chain
with $Z=112$. It may however be mentioned that for $Z=112$ the empirical data
available is limited to four data points. The number of data points 
available for $Z=114$ is limited to two points only. The trend of 
these two data points is represented quite well by the theoretical 
values, though the discrepancy for data point at $N=174$ is little 
more than 1 MeV. Given the complex picture of superheavy region 
that evolves due to interplay between deformation and single-particle 
states, the theoretical description of Q$_\alpha$ values as seen in
Figs.~\ref{fig:fig9} and ~\ref{fig:fig10} can be considered as
satisfactory.

\subsection{Alpha-decay half-lives}

Using the $Q_\alpha$-values, $\alpha$-decay half-lives can be calculated using
the phenomenological formula of Viola and Seaborg as given by,
\begin{equation}
\label{eqn:3.5}
log~T_\alpha=(aZ+b)Q_{\alpha}^{-1/2} +(cZ+d),
\end{equation}
where $Z$ is the atomic number of the parent nucleus and $Q_\alpha$ is the
$\alpha$-decay energy in MeV. Then, $T_\alpha$ is obtained in seconds. The
parameters $a, b, c$ and $d$ have been fitted to $\alpha$-decay half-lives of
known nuclei over a large range. This includes new experimental data 
obtained in the last decade. The values of $a$, $b$, $c$ and $d$ have 
been obtained by Sobiczewski et al. \cite{Sobicz.89}. 
The values of these parameter are 
$a=1.66175$, $b=-8.5166$, $c=-0.20228$, and $d=-33.9069$.

\begin{figure}[h!]
\begin{center}
\includegraphics[width=0.60\textwidth]{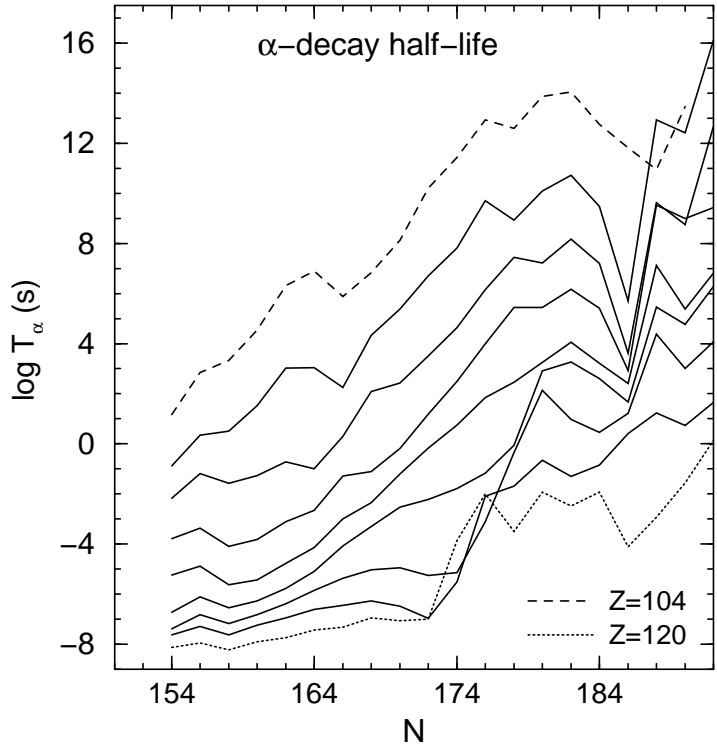}
\caption[T$_\alpha$ values with NL-SV1]{The $log~T_\alpha$ values  
for the isotopic chains with $Z=102-120$ and $N=150-190$ 
obtained with NL-SV1.}
\label{fig:fig11}
\end{center}
\end{figure}

Using the Viola and Seaborg formula of Eq. ~(\ref{eqn:3.5}), we have 
calculated the $\alpha$-decay half-lives $T_\alpha$ for nuclides in 
all the isotopic chains. The results are shown in 
Fig.~\ref{fig:fig11}. As can be seen from Eq. (\ref{eqn:3.5}),
$log~T_\alpha$ is inversely proportional to $\sqrt{Q_\alpha}$. The
decreasing trend of $Q_\alpha$ with neutron number as shown in
Fig.~\ref{fig:fig8} translates into an increasing trend 
for $log~T_\alpha$. As seen in Fig.~\ref{fig:fig11}, 
~$log~T_\alpha$ values show a rather linear increase with neutron number 
in the region $N=150$ to $N=180$ for all the isotopic chains with the exception
of the heaviest SHE $Z=116-120$. The latter show a linear 
behaviour in the curves in the neutron region $N=150-170$. 
Above $N=170$ the slope exhibits a further
increase in the value of $log~T_\alpha$. A nearly linear increase of
$log~T_\alpha$ with neutron number implies an exponential increase in
$T_\alpha$ in going to higher neutron numbers. The half-life curve shows
the highest values for the isotopic chain $Z=102$. As the $Z$ value of 
an isotopic chain increases, the half-life curves show a gradual decrease. 
This shows that in going to heavier SHE, nuclides become less and less 
unstable. 

Commensurate with Q$_\alpha$ values in Fig.~\ref{fig:fig8}, 
one notices a few peak-like structures in half-lives at certain 
neutron numbers. A small peak-like structure can be seen at 
the neutron number $N=154$ in nearly all the isotopic chains 
with the exception of the heaviest chains $Z=118$ and $Z=120$. 
The increased $T_\alpha$ at $N=154$ can be understood in 
terms of a deformed shell gap at this neutron
number. A similar behaviour (feature) has also been predicted by
non-relativistic macroscopic-microscopic calculations \cite{Moeller.94}.

Another minor enhancement in $T_\alpha$ can be seen in the vicinity of
$N \sim 160-162$ for the isotopic chains for the lightest SHE $Z=102-112$. 
The region of $N \sim 162$ has also been predicted to  be a deformed shell 
\cite{Moeller.94}. A major enhancement in $T_\alpha$ values 
can be conspicuously seen near $N=182-184$. A peak like structure in 
$log~T_\alpha$ curves can be seen near $N \sim 184$ for the isotopic chains 
of SHE with $Z=102-112$. Apparently, $N=182$ exhibits a higher value of
$T_\alpha$ as compared to $N=184$. In going to $N=186$, there 
is a sudden decrease in $T_\alpha$ values by a few orders of 
magnitude especially for lighter SHE with $Z=104-108$. For the chains with 
$Z=110-112$, there is also a decrease in $T_\alpha$ at $N=186$. This 
decrease is, however, smaller than for lighter SHE. Thus, for SHE 
with $Z=102-114$, the region of neutron number $N=184$ denotes 
a region of an enhanced stability against $\alpha$-decay. 

As one goes above $Z=114$ to heavier SHE, the $T_\alpha$ values 
show a significant decrease by several orders of magnitude. 
Nonetheless, SHE with $Z=116$ to $Z=120$ still show a 
maximum near $N\sim 184$. This implies that even for the 
heavier SHE the region of $N \sim 184$ shows a enhanced
stability as compared to their lighter neutron counterparts. Thus, the region
with $N=184$ would be suitable for synthesizing heavier superheavy
elements. For the case of $Z=120$ isotopic chain, there is also a peak in the
$T_\alpha$ curve in the Fig.~\ref{fig:fig11} at $N=174$. This indicates
that the combination $Z=120$ and $N=174$ may be suitable for synthesizing the
$Z=120$ superheavy element.

\subsection{Potential energy landscapes}

A potential energy landscape represents the total energy of a nucleus as a
function of deformation. It reflects how a nucleus rearranges its 
single-particle structure under the influence of deformation. 
We have carried out constrained RMF calculations with the 
Lagrangian set NL-SV1. The quadrupole deformation 
$\beta_2$ is used as a constraint parameter i.e, the binding energy of
a nucleus is calculated for a given deformation. The RMF calculations
are carried out for each value of $\beta_2$ in the deformation
plane with a small grid in $\beta_2$.

\begin{figure}[h!]
\begin{center}
\includegraphics[width=0.65\textwidth]{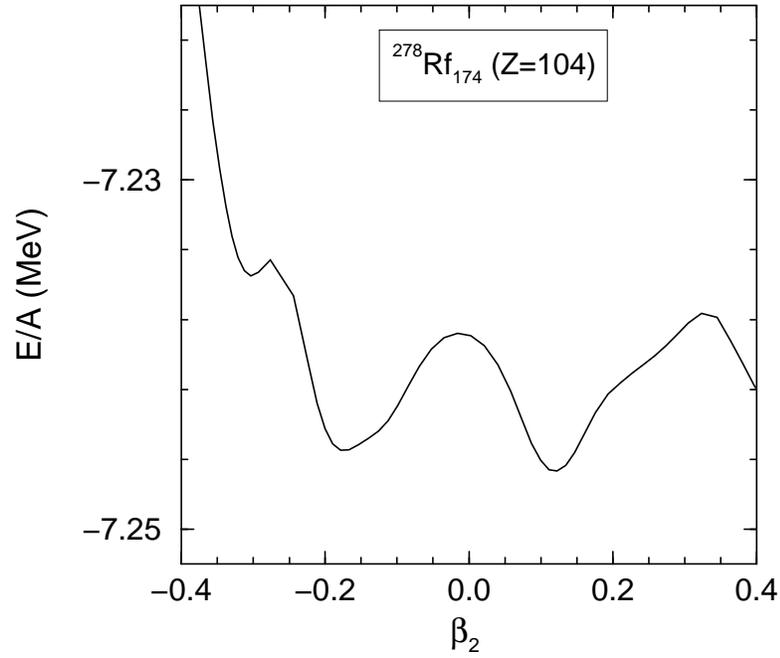}
\caption{The potential energy landscape for $^{278}$Rf$_{174}$ obtained with
  NL-SV1. }
\label{fig:fig12}
\end{center}
\end{figure}
\begin{figure}[h!]
\begin{center}
\includegraphics[width=0.65\textwidth]{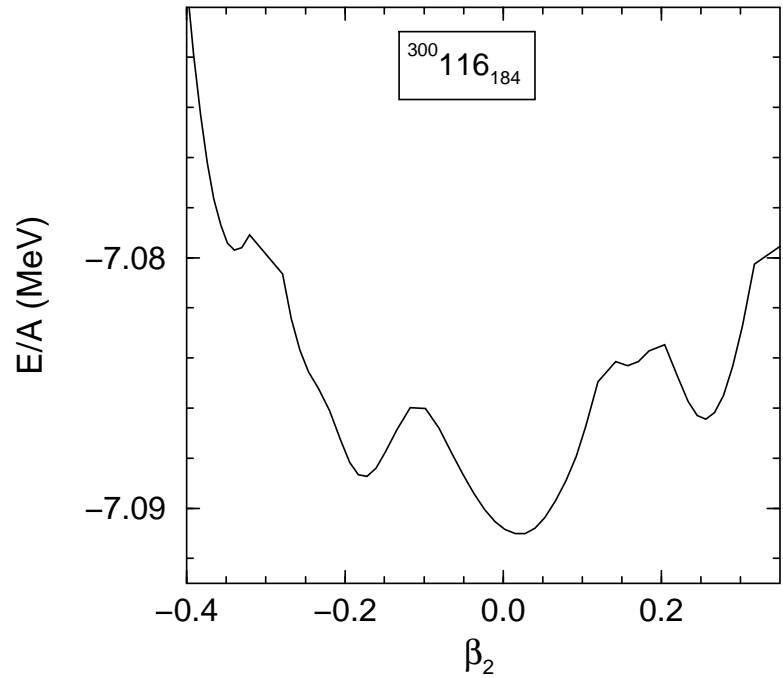}
\caption{The potential energy landscape for $^{300}116_{184}$ obtained with
  NL-SV1. }
\label{fig:fig13}
\end{center}
\end{figure}
\begin{figure}[h!]
\begin{center}
\includegraphics[width=0.65\textwidth]{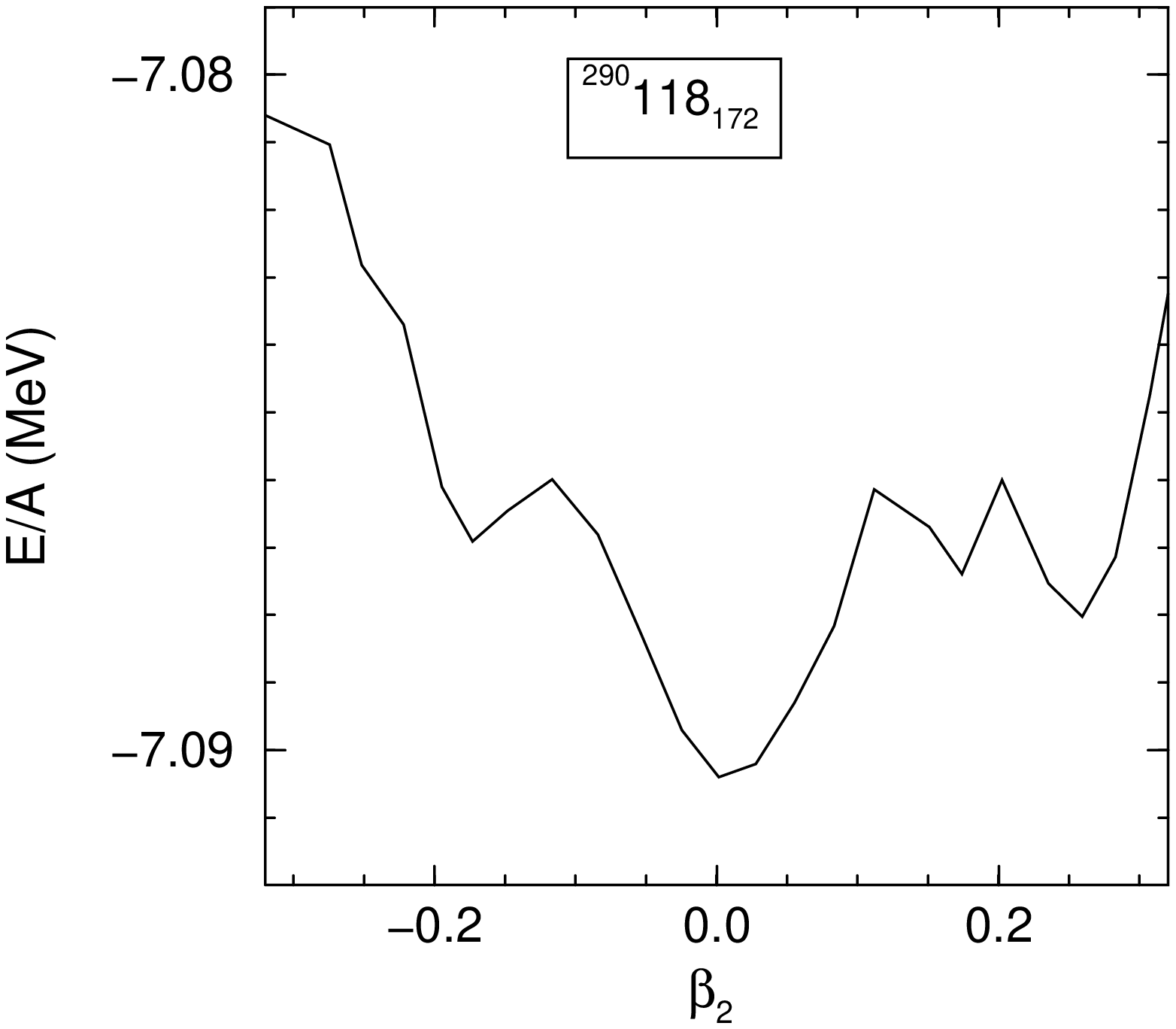}
\caption{The potential energy landscape for $^{290}118_{172}$ obtained with
  NL-SV1. }
\label{fig:fig14}
\end{center}
\end{figure}

We show in Fig.~\ref{fig:fig12} the potential energy surface (PES) 
for a nuclide of the lighter superheavy element  
$^{278}_{104}$Rf ($Z=104$) with the neutron 
number $N=174$. Two valleys can be seen, one in an oblate 
region and another in the prolate region. This picture is a clear 
reflection of a prolate-oblate shape co-existence whereby
the two minima are nearly degenerate in energy. 
The actual ground state in this case is, however, prolate as shown by 
the slightly lower valley in the prolate region. The PES for 
a heavier SHE for the nuclide
$^{300}116$ ($Z=116$) with the neutron number $N=184$ is shown in
Fig.~\ref{fig:fig13}. This figure illustrates a shape co-existence 
between a spherical and an oblate shape, as has been indicated by 
a blue square in Fig.~\ref{fig:fig5}. However, here the ground 
state is spherical in shape. This is consistent with $N=184$
still retaining some magicity in going to $Z=116$.

We show in Fig.~\ref{fig:fig14} the potential energy landscape for 
another nuclide with $Z=118$ and $N=172$. It shows a clear minimum 
for a spherical shape ($\beta_2$=0). There is no shape co-existence 
with any other shape for this nucleus. The spherical shape 
at $N=172$ reflects the character of a shell closure in the midst of 
deformed nuclei in a very heavy superheavy region. A spherical shape
for $N=172$ in the deformed region is accentuated by the 
proximity of its proton number to the presumed magic number $Z=120$.

\subsection{The single-particle levels}

The single-particle structure of nuclei is a good reflection of
magic numbers, whereby larger shell gaps at a magic number can be seen in
single-particle levels. In order to visualize some of the potential magic
numbers in the region of superheavy nuclei, here we show the single-particle
spectrum for a few key nuclei.
\begin{figure}[h!]
\begin{center}
\includegraphics[width=0.50\textwidth]{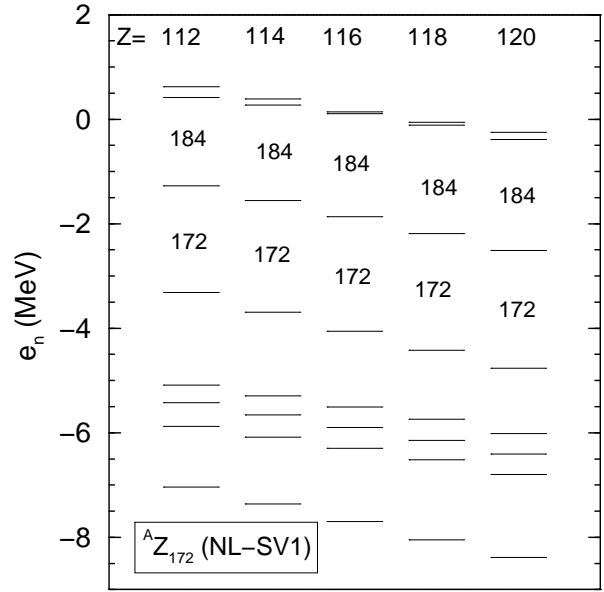}
\caption{The neutron single-particle levels for $N=172$ isotones for 
$Z$ = 112, 114, 116, 118, 120 with NL-SV1.}
\label{fig:fig15}
\end{center}
\end{figure}
As mentioned before, the neutron number $N=172$  is perceived as 
akin to a magic number. We show 
in Fig.~\ref{fig:fig15} neutron single-particle 
levels calculated for nuclides with $N=172$ for SHEs with $Z=112-120$. 
In all these nuclei the shell gaps at $N=172$  and at $N=184$ can be seen
clearly. In these nuclei the shell gap at $N=172$ is maintained in going
from $Z=112$ to $Z=120$. Quantitatively, there is no change as such in 
the shell gap at $N=172$ in going from $Z=112$ to $Z=120$. However, the 
gap itself is pushed down in energy in reaching $Z=120$. Thus, 
the neutron number $N=172$ would be helpful in synthesizing the 
superheavy elements with $Z=118$ and $Z=120$.

\begin{figure}[h!]
\begin{center}
\includegraphics[width=0.50\textwidth]{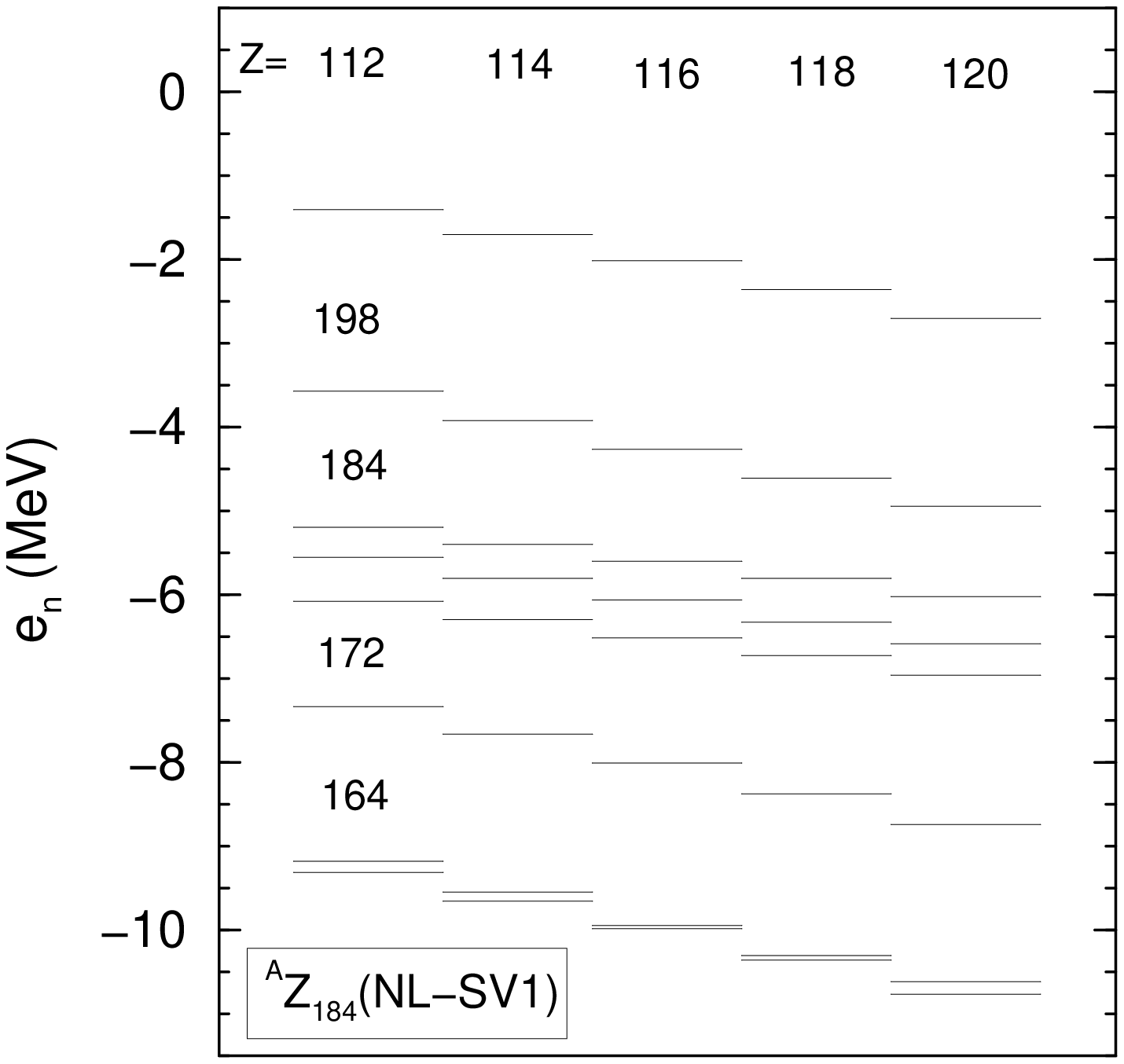}
\caption{The neutron single-particle energy levels for $Z$ = 112, 114, 116, 
118, 120 and  $N=184$ with NL-SV1.}
\label{fig:fig16}
\end{center}
\end{figure}
In Fig.~\ref{fig:fig16} we show the neutron single-particle 
spectrum for nuclides with $N=184$ for SHEs with $Z=112-120$. 
The shell gap at $N=184$ can be seen clearly. Thus, $N=184$
manifests as a major magic number. This is consistent with several other
theoretical predictions. In the RMF theory with NL-SV1, the shell gap at 
$N=184$ shows a decrease in going from $Z=112$ to $Z=120$. 
We also see a major shell gap at $N=198$. However, it may be of 
little practical importance in synthesis of superheavy elements 
due to very large number of neutrons involved.
\begin{figure}[h!]
\begin{center}
\includegraphics[width=0.50\textwidth]{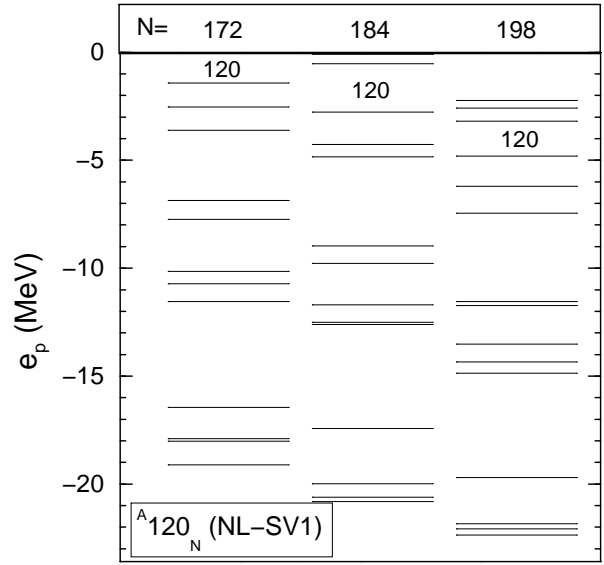}
\caption{The proton single-particle levels for the isotopes of $Z=120$ with 
$N$ = 172, 184 and 198 using NL-SV1.}
\label{fig:fig17}
\end{center}
\end{figure}

In order to visualize the single-particle structure of SHE $Z=120$, 
we show in Fig.~\ref{fig:fig17}, the proton single-particle levels obtained 
with NL-SV1 for nuclides with $N=172$, $N=184$ and $N=198$. The nuclide 
with $Z=120$ and $N=172$ is very close to being unbound. The proton shell 
gap at $Z=120$ can be seen for $N=172$, though it couples to continuum. 
From a practical point of view, the nuclide with $Z=120$ and $N=172$,
which shows the character of a doubly magic nucleus can be useful in
synthesizing SHE with $Z=118$ and $Z=120$. The combination $Z=120$ 
and $N=184$ also comes out as a doubly magic configuration. 
It may be possibly of use in synthesizing the heaviest superheavy
nuclei.

\section{Summary and Conclusions}

We have studied the structure of superheavy nuclei within the 
framework of the relativistic mean-field theory. The region of 
superheavy elements from $Z=102$ to $Z=120$ has been explored. 
The isotopic chains of the superheavy elements 
encompassing the neutron number from $N=150$ to $N=190$ have been 
investigated. The Lagrangian model NL-SV1 with the inclusion of the vector
self-coupling of the $\omega$-meson has been employed. 

RMF calculations for a large number of nuclides in the isotopic chains of
superheavy elements have been performed by taking an axially symmetric
deformed configuration. The binding energies, deformation properties 
and single-particle levels have been obtained from RMF+BCS minimizations. 
Nuclides in general are found to possess a moderate 
oblate or a prolate deformation. On the other hand, 
nuclei in the vicinity of the neutron number $N=184$ display a 
spherical shape.  Display of a spherical shape for all the isotopic
chains indicates  an inclination of $N=184$ being toward a potential magicity. 

It is shown that nuclei in the region of superheavy elements 
exhibit the phenomenon of shape co-existence. 
A large number of nuclei much below $N=184$ exhibit a shape co-existence 
between a prolate and an oblate shape. This is specially the case for 
the isotopic chains from $Z=102$ to $Z=114$. Some isotopic chains 
exhibit a shape-coexistence between a spherical and an oblate shape especially
those in the vicinity of $N=184$.  The shape-coexistence in nuclei is 
demonstrated by the potential energy landscapes obtained for a few nuclei. 

The $Q_\alpha$ values and the corresponding $\alpha$-decay half-lives were
obtained using the ground-state binding energies.
The $\alpha$-decay half-lives $T_\alpha$ demonstrate clearly that in the 
vicinity of $N=184$ nuclides have a significantly larger $T_\alpha$ as 
compared to their neighbours. This is again a strong indicator of the 
magicity of the neutron number $N=184$.

Two-neutron separation energies of nuclides were calculated using 
the results with NL-SV1. The $S_{2n}$ values exhibit a small kink
near $N=160$ and $N=172$. This suggests that the neutron number $N=160$ and
$N=172$ have a larger shell gap than their neighbours. In comparison, 
a strong kink in $S_{2n}$ values at $N=184$ is a clearest indication of a major
shell gap at $N=184$. 

The single-particle levels obtained with the Lagrangian 
model NL-SV1 show that there exist shell gaps at neutron numbers 
$N=172$ and $N=184$. Whilst $N=172$ can not be construed as a major magic 
number as displayed by a mild kink in $S_{2n}$ values, 
the neutron number $N=184$ does behave as a major magic number in the 
neutron-rich region. Evidently, the magicity of $N=184$ is demonstrated 
succintly by a large shell gap at $N=184$ seen in the single-particle levels. 
This picture of the magicity of N=184 in the RMF theory is consistent 
with the predictions in the density-dependent Skyrme Hartree-Fock approach and
also with the previous studies performed within the RMF theory using the
conventional model of the nonlinear scalar self-couplings. 
Thus, our study with the Lagrangian model NL-SV1 with the vector 
self-coupling of $\omega$ meson reinforces the magic character of the 
neutron number $N=184$. Albeit, a decrease in the shell gap at $N=184$ 
in going towards $Z=120$ implies a weakening of the shell strength in 
approaching the extreme end of the periodic table. 

Whilst a combination of $N=184$ in conjuction with $Z$ in the vicinity of
120 produces nuclei which have impractically large number of neutrons,
the combination of the neutron number $N=172$ and $Z=120$ does lie within
the domain of feasibility. Our study shows that this combination 
is suggestive of a double magicity. Thus, synthesis of superheavy 
elements heavier than $Z=116$ with neutron number close to $N=172$ 
should be susceptible to an extra stability rendered by 
the apparent magic nature of $Z=120$ and $N=172$.\\ 

\section{Acknowledgments}
This work is supported by the Graduate Research Project Y01/05 
of the Research Administration, Kuwait University. Useful discussions
with Professor G. M\"unzenberg are thankfully acknowledged.

\end{document}